\RequirePackage[T1]{fontenc}
\documentclass[12pt]{article}

\usepackage[height=8.85in,width=6.25in]{geometry}

\usepackage[utf8]{inputenc}
\usepackage{amsmath}
\usepackage{amssymb}
\usepackage{mathtools}
\numberwithin{equation}{section}
\usepackage{slashed}
\usepackage{braket}
\usepackage[svgnames,dvipsnames]{xcolor}
\usepackage[colorlinks,citecolor=DarkGreen,linkcolor=FireBrick,urlcolor=FireBrick,linktocpage,breaklinks=true]{hyperref}
\usepackage{cite}
\usepackage{graphicx}
\usepackage{tikz}
\usetikzlibrary{arrows.meta,decorations.markings}
\tikzset{
    line/.style={thick,postaction={decorate},decoration={markings,
        mark=at position 0.55 with {\arrow{Stealth[length=1.8mm]}}}},
    sline/.style={thick,postaction={decorate},decoration={markings,
        mark=at position 0.6 with {\arrow{Stealth[length=1.4mm]}}}},
}
\newcommand{\fourtreecoords}{%
    \coordinate (L1) at (-1.2,-1); \coordinate (L2) at (-0.4,-1);
    \coordinate (L3) at (0.4,-1);  \coordinate (L4) at (1.2,-1);
    \coordinate (N12) at (-0.8,-0.52); \coordinate (N23) at (0,-0.52);
    \coordinate (N34) at (0.8,-0.52);
    \coordinate (N13) at (-0.4,-0.04); \coordinate (N24) at (0.4,-0.04);
    \coordinate (R) at (0,0.44); \coordinate (T) at (0,1);
    \node[below] at (L1) {$X$}; \node[below] at (L2) {$Y$};
    \node[below] at (L3) {$Z$}; \node[below] at (L4) {$W$};
    \draw[sline] (R) -- (T);}
\tikzset{blob/.style={circle,fill,inner sep=0pt,minimum size=3.6pt}}
\newcommand{\treeblobs}[1]{\foreach \p in {#1} {\node[blob] at (\p) {};}}

\usepackage{times}
\usepackage{courier}
\usepackage{bm}
\usepackage{subfig}
\usepackage{here}

\DeclareMathOperator{\Hom}{Hom}
\DeclareMathOperator{\End}{End}
\DeclareMathOperator{\Cl}{Cl}
\DeclareMathOperator{\id}{id}
\def\bZ{\mathbb{Z}}
\def\bR{\mathbb{R}}
\def\bF{\mathbb{F}}
\def\bC{\mathbb{C}}
\def\cA{\mathcal{A}}
\def\cC{\mathcal{C}}
\def\cD{\mathcal{D}}
\def\cR{\mathcal{R}}

\begin{document}

\begin{titlepage}

\begin{flushright}

\end{flushright}

\vskip 3cm

\begin{center}

{\Large \bfseries Fermionic  anomalies in 1+1 dimensions \\[1em]
from a categorical point of view}

\vskip 1cm
Kotaro Kawasumi and  Yuji Tachikawa
\vskip 1cm

\begin{tabular}{ll}
& Kavli Institute for the Physics and Mathematics of the Universe (WPI), \\
& University of Tokyo,  Kashiwa, Chiba 277-8583, Japan
\end{tabular}

\vskip 2cm

\end{center}

\noindent 
One way to understand the anomaly of finite group symmetries of bosonic systems in 1+1 dimensions is 
to consider fusion categories whose simple objects are all invertible.
We develop an analogous description for the fermionic case 
using super fusion categories.
It turns out that, due to the possible existence of Majorana zero modes, 
we need to slightly weaken the condition that all simple objects are invertible.
We show that the analysis of super fusion categories where
all simple objects are `weakly invertible' perfectly reproduces
the known classification of fermionic anomalies in 1+1 dimensions
using spin bordism groups.

Along the way, we also review the relation between symmetries
of quantum systems in 1+1 dimensions and fusion categories,
both from the point of view of topological line operators
and from the point of view of endomorphisms of the algebra of local operators.
We will provide explicit examples of the latter constructions in the appendices.

\end{titlepage}

\setcounter{tocdepth}{2}
\tableofcontents

\section{Introduction}

In the last decade it was firmly established that the anomaly of finite group symmetry of a $(d+1)$-dimensional system is encoded 
in terms of a $(d+2)$-dimensional symmetry-protected topological (SPT) phase,
which is in turn classified using an appropriate bordism group \cite{Kapustin:2014dxa,Freed:2016rqq,Yonekura:2018ufj}.
This general statement grew out of the study of anomalies and SPT phases
by many researchers, whose original impetus came  from the work of Gu, Wen and collaborators, e.g.~\cite{Chen:2011pg,Gu:2012ib}.
For example, the anomaly of a finite group symmetry $G$ in 1+1 dimensions
is given by $H^3(BG;U(1))$ in the bosonic case, and by 
$\Hom(\tilde\Omega_3^{\text{spin}}(BG),U(1))$ in the fermionic case,
the structure of which was determined in \cite{BrumfielMorgan1}.

The same issue can be studied also from a categorical point of view,
by studying how various topological defects implementing the symmetry actions fuse with each other \cite{Bhardwaj:2017xup,Chang:2018iay}.
In the case of bosonic systems in 1+1 dimensions, the topological defect lines form fusion categories, in which the symmetry actions are not necessarily invertible.
Ordinary finite group symmetries correspond to the special situation
where all symmetry actions are invertible,
corresponding to fusion categories whose simple objects are all invertible.
Such fusion categories can be easily classified, and
are indeed characterized by the data of a finite group $G$ and a class in $H^3(BG;U(1))$, which is an elementary result in the theory of fusion categories \cite{EGNO}.
This reproduces the known analysis from the geometric study of SPT phases.

The aim of this paper is to extend this categorical analysis to the fermionic case, 
and to reproduce the known classification of fermionic anomalies in 1+1 dimensions using spin bordism groups.
The main new ingredient in the fermionic case is that
the topological defect lines implementing the symmetry actions can have
Majorana fermion zero modes even in the group symmetry case,
making the fusion of two such lines not quite invertible but only `weakly' so,
in a sense which we will make precise.
The authors are aware that most of the ingredients are already in the literature, e.g.~in \cite{Else:2014vma,Cheng:2015yoa,Seifnashri:2023dpa},
but it is their opinion that nontrivial steps and efforts were required to fill in the omitted details.
This is why this paper is written, and 
the authors hope that the details it contains will be useful to the readers.

The essential idea is the following.
From a categorical point of view, the main difference of fermionic symmetries 
from bosonic ones is that they are described by super fusion categories,\footnote{%
This point has been discussed in many papers, not all of which can be cited here.
See e.g.~\cite{Inamura:2022lun,Bhardwaj:2024ydc,Chang:2026pdu} for a sample of recent papers.
}
that is,  the Hom space $\Hom(X,Y)$ between two objects $X$ and $Y$ 
can be split into the bosonic and fermionic parts:\[
\Hom(X,Y)=\Hom(X,Y)_0\oplus \Hom(X,Y)_1.
\] Equivalently, the Hom space $\Hom(X,Y)$ is a super vector space.
Then, the endomorphism space $\End(X):=\Hom(X,X)$ of an object $X$ 
is a superalgebra. 
An object $X$ is simple when $\End(X)$ does not contain any nontrivial 
bosonic projection operator that can be used to decompose it further.
This means that $\End(X)$ is either $\bC$ or 
the Clifford algebra generated by a single odd element $\psi$,
$\Cl(\psi)=\bC\oplus \bC\psi$.
The operator $\psi\in \End(X)$ is the Majorana zero mode on the topological defect line $X$.

In the bosonic case, a group symmetry operation corresponds to 
an invertible simple object $X$, that is, we have a dual object $X^\vee$ such that 
\begin{equation}
X\otimes X^\vee\simeq 1. \label{eq:invertible}
\end{equation}
But this never happens when $\End(X)\simeq \Cl(\psi)$, and we need to weaken the notion of invertibility.
The minimal weakening in such a case is to require that 
\begin{equation}
X\otimes X^\vee\simeq 1\oplus \Pi. \label{eq:weakly-invertible}
\end{equation}
Here, $\Pi$ is a simple object almost identical to $1$,
but such that the Hom space $\Hom(1,\Pi)$ is 
purely fermionic and one-dimensional.
This object $\Pi$ is useful in keeping track of the fermion parity of the system.
Then the $\End$ space of both sides of \eqref{eq:weakly-invertible} 
is isomorphic to the Clifford algebra with two generators $\Cl(\psi_1,\psi_2)$.

We call an object $X$ satisfying \eqref{eq:weakly-invertible} `weakly invertible',
with the understanding that ordinary invertible objects are also considered weakly invertible.
Our main result is to show that the super fusion categories where all simple objects are weakly invertible are classified by the data of a finite group $G$,
together with a triple of group cochains satisfying certain conditions,
which reproduces the explicit description of $\Hom(\tilde\Omega_3^{\text{spin}}(BG),U(1))$ given in \cite{BrumfielMorgan1}.
Extracting the relevant cochains from a given such super fusion category 
requires what we call a `coordinate system' and some use of basic results
in the theory of Clifford algebras,
which was not required in the bosonic case, where every Hom space involved 
is a one-dimensional vector space.
This is the most nontrivial step in the analysis, which we will detail below.

This paper is intended for readers who study high energy physics theory
or condensed matter physics theory, with a basic background in the theory of fusion categories,
but not intended for rigorous mathematicians. 
For them, one of the authors (KK) has a paper \cite{Kawasumi},
where a mathematically rigorous treatment of the same subject
and much more will be given, working not only over $\bC$ but also over other fields, including $\bR$ and $\bF_q$,
and also providing a Lean formalization.
In a sense the present paper is only a subset of \cite{Kawasumi}, but 
it is hoped that this will be more accessible to physicists.

The rest of the paper is organized as follows:
\begin{itemize}
\item We start in Sec.~\ref{sec:review} by briefly reviewing the relation
between symmetries of quantum systems in 1+1 dimensions and fusion categories.
We will employ two different perspectives, one from the point of view of topological line operators, and the other from the point of view of endomorphisms of the algebra of local operators.
The former is more commonly used in the hep-th community,
while the latter is closer to the approach more commonly used in the cond-mat community.
The latter approach is also the one which is usually used in the algebraic quantum field theory community. 
\item 
In Sec.~\ref{sec:bosonic}, as a warm-up, we recall
the basic case of anomalies of finite group symmetries
of bosonic systems in 1+1 dimensions.
Fusion categories whose simple objects are all invertible are usually called
pointed fusion categories, and we review their classification in terms of a finite group $G$ and a class in $H^3(BG;U(1))$.
\item 
Then, in Sec.~\ref{sec:fermionicI} we study super fusion categories whose simple objects are all invertible, which we call pointed super fusion categories.
We show that they are classified by a finite group $G$ and a class in
the supercohomology group $SH^3(BG)$, first introduced by Gu and Wen in \cite{Gu:2012ib}.
The analysis in this case can be done as a straightforward generalization of the bosonic case.
\item 
The main nontrivial part of the paper is Sec.~\ref{sec:fermionicII}, where
we study super fusion categories whose simple objects are all weakly invertible, which we call weakly-pointed super fusion categories.
The analysis is significantly more involved,
but in the end we will find a result perfectly compatible
with the known classification of fermionic anomalies in 1+1 dimensions using spin bordism groups.
\item We conclude the paper in Sec.~\ref{sec:comparison} by checking
the agreement of the outcome of our categorical analysis and 
that of the geometric analysis of \cite{BrumfielMorgan1}.
\item We also have two appendices, where we construct explicit examples of
categories from the endomorphisms of the algebra of local operators on spin chains.
In Appendix~\ref{app:spinchain}, we consider an explicit spin chain
with an arbitrary finite group symmetry and an anomaly $\omega\in H^3(BG;U(1))$,
based on the group-cohomology models of SPT phases \cite{Chen:2011pg},
and construct a pointed fusion category from the endomorphisms of the algebra of local operators of this spin chain.
This can be considered a translation of the method of Else and Nayak in \cite[Appendix~E]{Else:2014vma} to this language.
Then, in Appendix~\ref{app:reference}, we construct a `reference' weakly-pointed
super fusion category, corresponding to a $\bZ_2$ symmetry with a Majorana zero mode, 
using a fermionic spin chain and an associated infinite-dimensional algebra.
This reference category will be needed in the analysis of Sec.~\ref{sec:fermionicII}.
\end{itemize}

\section{Symmetry and categories: a very brief review}
\label{sec:review}
Let us start with a brief review of the relation between symmetries of
quantum systems in 1+1 dimensions and fusion categories.
We will review it from two perspectives,
one from the point of view of the topological line operators,
and the other from the point of view of automorphisms and 
endomorphisms of the algebra of local operators.
The former is more commonly used in the hep-th community,
while the latter is the approach typically used in the algebraic quantum field theory community.
The approach used in the condensed matter theory community is not
directly equal to the latter approach but is closely related to it.
Therefore, the authors feel it appropriate to review both points of view.

\subsection{Symmetry and the topological lines}
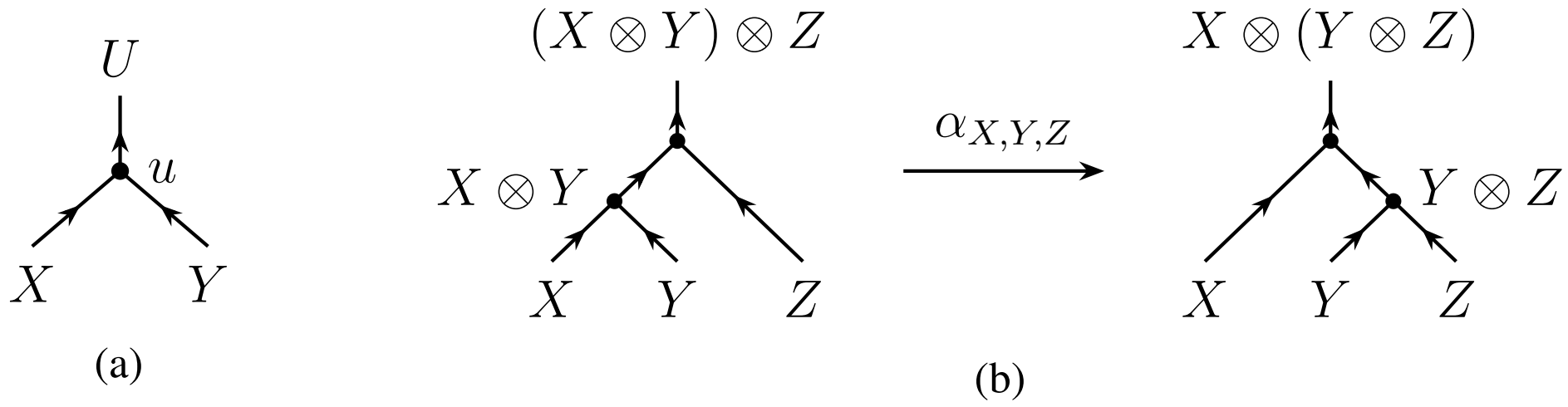
\begin{figure}[h]
    \centering
    \subfloat[\label{fig:junction-a}]{%
    \begin{tikzpicture}[baseline=(u),yscale=.6]
        \coordinate (u) at (0,0);
        \draw[line] (-0.7,-1) node[below] {$X$} -- (u);
        \draw[line] (0.7,-1) node[below] {$Y$} -- (u);
        \draw[line] (u) -- (0,1) node[above] {$U$};
        \node[blob,minimum size=4pt] at (u) {};
        \node[right=2pt] at (u) {$u$};
    \end{tikzpicture}}
    \hspace{3em}
    \subfloat[\label{fig:junction-b}]{%
    \begin{tikzpicture}[baseline=(m),yscale=.6]
        \coordinate (m) at (0,0);
        \begin{scope}[xshift=-2.6cm]
            \coordinate (a) at (-0.5,-0.4);
            \coordinate (b) at (0,0.4);
            \draw[line] (-1,-1.2) node[below] {$X$} -- (a);
            \draw[line] (0,-1.2) node[below] {$Y$} -- (a);
            \draw[line] (a) -- node[pos=0.2,left=4pt] {$X\otimes Y$} (b);
            \draw[line] (1,-1.2) node[below] {$Z$} -- (b);
            \draw[line] (b) -- (0,1.2) node[above] {$(X\otimes Y)\otimes Z$};
            \treeblobs{a,b}
        \end{scope}
        \begin{scope}[xshift=2.6cm]
            \coordinate (a) at (0.5,-0.4);
            \coordinate (b) at (0,0.4);
            \draw[line] (-1,-1.2) node[below] {$X$} -- (b);
            \draw[line] (0,-1.2) node[below] {$Y$} -- (a);
            \draw[line] (1,-1.2) node[below] {$Z$} -- (a);
            \draw[line] (a) -- node[pos=0.2,right=4pt] {$Y\otimes Z$} (b);
            \draw[line] (b) -- (0,1.2) node[above] {$X\otimes(Y\otimes Z)$};
            \treeblobs{a,b}
        \end{scope}
        \draw[-{Stealth[length=2mm]},thick] (-0.8,0) -- node[above] {$\alpha_{X,Y,Z}$} (0.8,0);
    \end{tikzpicture}}
    \caption{(a) Two topological lines $X$ and $Y$ fuse into $U$ at a junction,
    where we specify a local operator $u\in\Hom(X\otimes Y,U)$.
    (b) The associator $\alpha_{X,Y,Z}$ relates the two orders of fusing three lines.
    }
    \label{fig:junction}
\end{figure}

When we use a spacetime picture to describe a quantum system in 1+1 dimensions,
a symmetry operation $X$ can be thought of as a topological line operator
in the spacetime, across which the system experiences the said symmetry operation.
If we have two such symmetry operations $X$ and $Y$
represented by two topological lines,
we can fuse them together to form a new topological line, say $U$.
There can be more than one way to do so.
To distinguish them, we label the junction with a local operator $u$
encoding the way we connect the lines.
We denote the space of such local operators by $\Hom(X\otimes Y,U)$,
that is, $u\in \Hom(X\otimes Y,U)$. See Fig.~\ref{fig:junction}
for an illustration.

When the symmetry is described by a finite group $G$,
we will have lines $X_g$ and $X_h$ for $g,h\in G$.
Then the fusion of $X_g$ and $X_h$ should give a unique line, which is $X_{gh}$.
More generally, there can be multiple topological lines $U$ emerging from the fusion of $X$ and $Y$, giving the structure \begin{equation}
X\otimes Y\simeq \bigoplus_i U_i.
\end{equation}
These two cases are often distinguished by saying that the former and the latter 
correspond to invertible and non-invertible symmetries, respectively.
Namely, in the former, given $X_g$, we have the inverse operation $X_{g^{-1}}$ such that
\begin{equation}
X_g\otimes X_{g^{-1}}\simeq 1,
\end{equation}
where $1$ is the trivial line.
In the latter, for any simple line $X$, there is a line with a reverse orientation, which we denote by $X^\vee$. 
Then $X\otimes X^\vee$ contains the trivial line $1$, but may contain more:
\begin{equation}
X\otimes X^\vee\simeq 1\oplus\cdots .
\end{equation}
Therefore, $X$ has what can be called a dual $X^\vee$,
but their composition gives not just the identity, but some more.
One important point in our paper is that when we consider fermionic systems,
even a group symmetry requires us to consider `slightly non-invertible' 
topological lines, as we will see below.

Let us now consider three topological lines $X$, $Y$ and $Z$.
When we do the fusion of the three lines, 
we can do it either in the order $(X\otimes Y)\otimes Z$ or in the order $X\otimes (Y\otimes Z)$.
There should be a way to relate them,
which is known as the associator $\alpha_{X,Y,Z}$:
\begin{equation}
\alpha_{X,Y,Z}: (X\otimes Y)\otimes Z\xrightarrow{\sim} X\otimes (Y\otimes Z).
\end{equation}
This is also pictured in Fig.~\ref{fig:junction}.

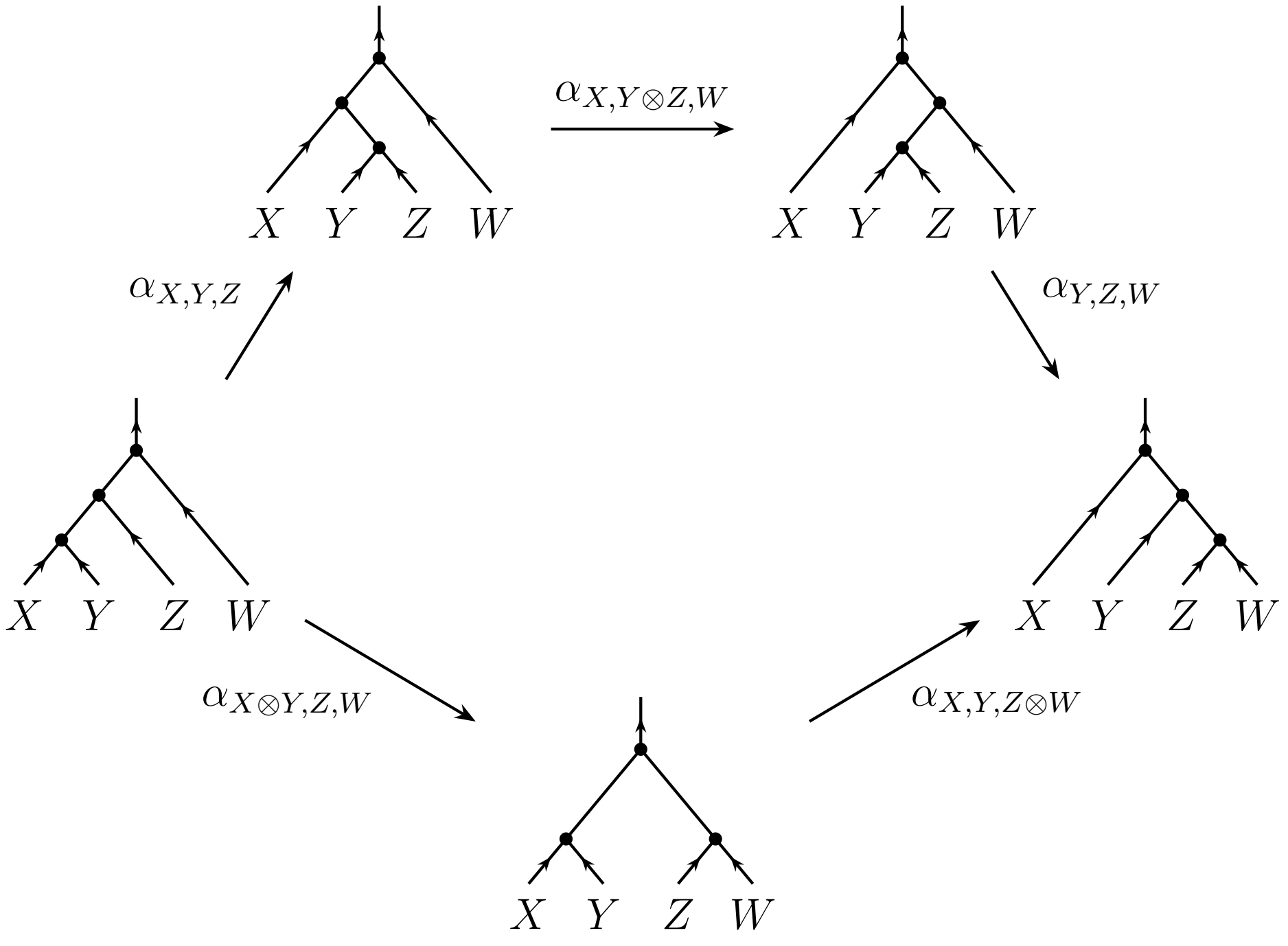
\begin{figure}
    \centering
    \begin{tikzpicture}[scale=0.9]
        \begin{scope}[shift={(-5.4,0)},local bounding box=A]
            \fourtreecoords
            \draw[sline] (L1) -- (N12); \draw[sline] (L2) -- (N12);
            \draw[thick] (N12) -- (N13); \draw[sline] (L3) -- (N13);
            \draw[thick] (N13) -- (R); \draw[sline] (L4) -- (R);
            \treeblobs{N12,N13,R}
        \end{scope}
        \begin{scope}[shift={(-2.8,4.2)},local bounding box=B]
            \fourtreecoords
            \draw[sline] (L2) -- (N23); \draw[sline] (L3) -- (N23);
            \draw[sline] (L1) -- (N13); \draw[thick] (N23) -- (N13);
            \draw[thick] (N13) -- (R); \draw[sline] (L4) -- (R);
            \treeblobs{N23,N13,R}
        \end{scope}
        \begin{scope}[shift={(2.8,4.2)},local bounding box=C]
            \fourtreecoords
            \draw[sline] (L2) -- (N23); \draw[sline] (L3) -- (N23);
            \draw[thick] (N23) -- (N24); \draw[sline] (L4) -- (N24);
            \draw[sline] (L1) -- (R); \draw[thick] (N24) -- (R);
            \treeblobs{N23,N24,R}
        \end{scope}
        \begin{scope}[shift={(0,-3.2)},local bounding box=D]
            \fourtreecoords
            \draw[sline] (L1) -- (N12); \draw[sline] (L2) -- (N12);
            \draw[sline] (L3) -- (N34); \draw[sline] (L4) -- (N34);
            \draw[thick] (N12) -- (R); \draw[thick] (N34) -- (R);
            \treeblobs{N12,N34,R}
        \end{scope}
        \begin{scope}[shift={(5.4,0)},local bounding box=E]
            \fourtreecoords
            \draw[sline] (L3) -- (N34); \draw[sline] (L4) -- (N34);
            \draw[sline] (L2) -- (N24); \draw[thick] (N34) -- (N24);
            \draw[sline] (L1) -- (R); \draw[thick] (N24) -- (R);
            \treeblobs{N34,N24,R}
        \end{scope}
        \begin{scope}[-{Stealth[length=2mm]},thick,shorten >=6pt,shorten <=6pt]
            \draw (A) -- node[above left] {$\alpha_{X,Y,Z}$} (B);
            \draw (B) -- node[above] {$\alpha_{X,Y\otimes Z,W}$} (C);
            \draw (C) -- node[above right] {$\alpha_{Y,Z,W}$} (E);
            \draw (A) -- node[below left] {$\alpha_{X\otimes Y,Z,W}$} (D);
            \draw (D) -- node[below right] {$\alpha_{X,Y,Z\otimes W}$} (E);
        \end{scope}
    \end{tikzpicture}
    \caption{The pentagon relation equates two ways of rearranging
    $((X\otimes Y)\otimes Z)\otimes W$ into $X\otimes(Y\otimes(Z\otimes W))$.}
    \label{fig:pentagon}
\end{figure}

When there are four topological lines $X$, $Y$, $Z$ and $W$ involved,
we can rearrange $((X\otimes Y)\otimes Z)\otimes W$ to $X\otimes (Y\otimes (Z\otimes W))$ in two different ways,
which need to agree. In an equation, this requirement is 
\begin{equation}
\alpha_{Y,Z,W}\circ \alpha_{X,Y\otimes Z,W}\circ \alpha_{X,Y,Z}
=
\alpha_{X,Y,Z\otimes W}\circ \alpha_{X\otimes Y,Z,W},
\end{equation}
while in pictures it becomes the diagram in Fig.~\ref{fig:pentagon}.
As the picture is a pentagon, this is known as the pentagon equation.

So far we recalled that topological lines $X,Y,Z,\ldots$ of quantum systems in 1+1 dimensions
can be fused together $X\otimes Y$, have duals $X^\vee,Y^\vee,Z^\vee,\ldots$, and have associators $\alpha_{X,Y,Z}$ satisfying the pentagon equation.
These structures are summarized by saying that
the topological lines form a fusion category.
This was pointed out in the 2010s in the high-energy theory community, see e.g.~\cite{Carqueville:2012dk,Bhardwaj:2017xup,Chang:2018iay}.

\subsection{Symmetry and the endomorphisms of the algebra of local operators}
\label{sec:symmetry-algebra}
Let us now consider the quantum system in 1+1 dimensions 
from a Hamiltonian perspective.
In the condensed matter community, the following analysis originates 
from the work  \cite{Else:2014vma} on quantum spin chains, again in the 2010s.  
For this reason, this method is often referred to as the Else-Nayak method.
The papers developing this method further are too numerous to be all cited here,
but see e.g.~\cite{Seifnashri:2023dpa,Chavda:2026eeq}.

It turns out that, in the context of algebraic quantum field theory,  essentially the same analysis was known well before Else and Nayak, 
ultimately going back to Longo's analysis \cite{Longo:1989tt} in the late 1980s.
For more recent accounts, see e.g.~\cite[Sec.~4.2]{Muger:2005ra} in the invertible case, and \cite{Bischoff:2014xea} for an extensive review.
Here we will only scratch the surface.

In this section, we will ignore the subtleties due to having
fermionic degrees of freedom.
These points will be treated later, starting in Sec.~\ref{sec:fermionicI}.

\subsubsection{Half-space symmetry actions on the algebra of local operators}
We put the system on an infinite line we parameterize by $x\in \bR$,
and consider the algebra $\cA$ of local operators around $x\sim 0$.
We consider performing a symmetry operation labeled by $X$ only on the $x>0$ half
of the system.
In the spacetime language of the previous subsection,
this operation creates a topological line labeled by $X$ at $x=0$,
extending in the temporal direction.
This operation does something to the algebra $\cA$:
\begin{equation}
\rho_X:\cA\to \cA.
\end{equation}
As $X$ is supposed to be a symmetry, 
we demand that it preserves essential features of $\cA$, i.e.~the product and the unit of $\cA$.
It is called an automorphism when $\rho_X(\cA)=\cA$, and an endomorphism when $\rho_X(\cA)\subset \cA$.
The second option, when the image is a proper subset,
can clearly only happen when $\cA$ is infinite-dimensional.\footnote{
    In the continuum, the algebra of local operators supported on a finite region will already be infinite-dimensional.
    In the spin chain case, one needs to consider the algebra of operators 
    which affect only finitely many sites.
    Even though each element of the algebra only affects finitely many sites, the algebra itself is infinite-dimensional.
}

\begin{figure}[h]
    \centering
    \begin{tikzpicture}[scale=1.2,yscale=0.6,
        declare function={step(\t)=3*\t^2-2*\t^3;}]
        \def\w{0.75}
        \fill[yellow!30] (0,-1) rectangle (\w,3.5);
        \draw (0,0) -- (2.7,0);
        \foreach \h/\c/\l in {1.75/red/X_h, 2.35/blue/X_g, 2.95/ForestGreen/X_{gh}} {
            \draw[thick,\c] plot[domain=0:\w,samples=40,smooth]
                (\x,{\h*step(\x/\w)}) -- (2.7,\h) node[right] {$\l$};
        }
        \draw[very thick,DarkRed] ({\w/2-2.95},0) -- (0,0);
        \node[below=3pt] at (0.85,0) {$u_{g,h}\in\cA$};
        \pgfresetboundingbox
        \path[use as bounding box] ({\w/2-2.95},-1) rectangle ({\w/2+2.95},3.5);
    \end{tikzpicture}
    \caption{Symmetry operations on the half line $x>0$.
    Composition of $X_g$ and $X_h$ and the action of $X_{gh}$
    can differ by a unitary $u_{g,h}\in\cA$ localized around $x\sim 0$.}
    \label{fig:half-line-action}
\end{figure}
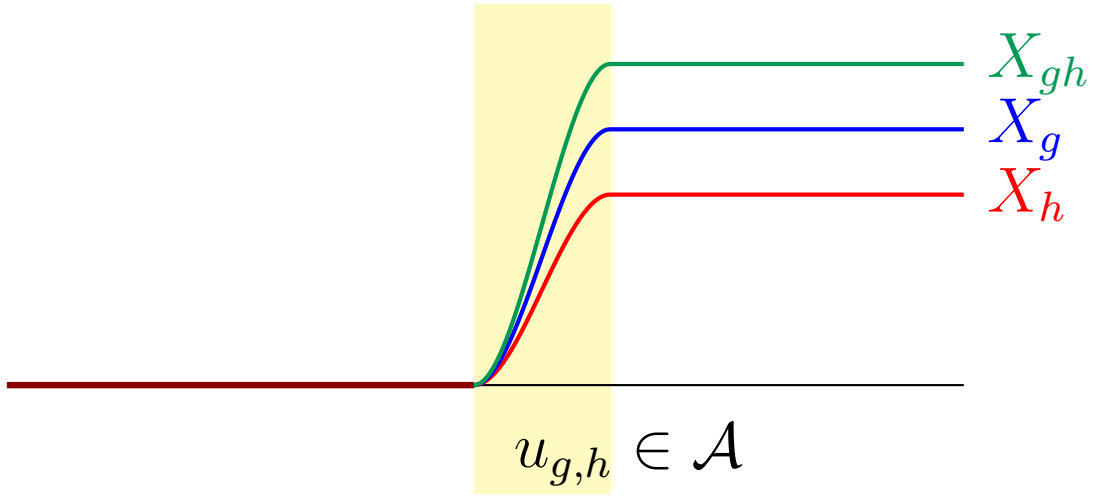

Consider the case of a group symmetry $G$. Let $g,h\in G$ be two elements.
Denote the symmetry operations on the half line $x>0$ by $X_g$ and $X_h$,
and   the corresponding endomorphisms of $\cA$ by $\rho_g$ and $\rho_h$.
Now, due to the non-trivial nature of the half-line symmetry action,
composition of $X_g$ and $X_h$ for $g,h\in G$ 
can differ from the action of $X_{gh}$ by some local disturbance
around the region $x\sim 0$,
which is represented by a unitary operator $u_{g,h}$ in the 
local algebra $\cA$. We will then have
\begin{equation}
\rho_{gh}(a)=u_{g,h}\rho_g(\rho_h(a))u_{g,h}^{-1},\qquad a\in \cA.
\label{eq:local-unitary}
\end{equation}
This is illustrated in Fig.~\ref{fig:half-line-action}.

This point of view was originally put forward in the condensed matter theory community
by Else and Nayak \cite[Appendix~E]{Else:2014vma}
in a slightly different formulation.
A recent paper \cite{Okada:2025kie} contains 
an explicit construction of $u_{g,h}$ for the cyclic group $\bZ_n$
embedded in a continuous $U(1)$ symmetry,
as an exponential of a local integral of the $U(1)$ current
in the region $x\sim 0$.
We also present a general construction of an infinite-dimensional algebra
whose endomorphisms realize a given finite group symmetry with a given anomaly in Appendix~\ref{app:spinchain}.

\subsubsection{Categorical interpretation of half-space symmetry actions}

Let us now give a categorical interpretation of the above construction.
Given an algebra $\cA$, we consider a category $\End(\cA)$, whose objects
are endomorphisms $\rho:\cA\to \cA$ of $\cA$.
The elements of $\Hom(\rho,\sigma)$ between two objects $\rho$ and $\sigma$ 
are declared to be their intertwiners $t\in \cA$,
that is, elements $t$ which satisfy the condition \begin{equation}
t\rho(a)=\sigma(a)t,\qquad  a\in \cA.
\end{equation}
The composition of two morphisms $t\in \Hom(\rho,\sigma)$ and $s\in \Hom(\sigma,\tau)$ is given simply by the product $st\in \Hom(\rho,\tau)$ in $\cA$.

Note that the local unitary condition \eqref{eq:local-unitary} we saw above
can be rewritten as \begin{equation}
u_{g,h}\rho_g(\rho_h(a))=\rho_{gh}(a)u_{g,h},\qquad a\in \cA,
\end{equation} that is, \begin{equation}
u_{g,h}\in \Hom(\rho_g\circ \rho_h,\rho_{gh})
\end{equation} is an intertwiner from $\rho_g\circ \rho_h$ to $\rho_{gh}$.
But note that we do not require a general intertwiner to be unitary,
and the space of intertwiners is a linear space.

This motivates us to define the tensor product in this category to be simply
the composition of endomorphisms: \begin{equation}
\rho\otimes \sigma := \rho\circ \sigma.
\end{equation}
The associator is then trivial, since $\rho\circ (\sigma\circ \tau) = (\rho\circ \sigma)\circ \tau$,
and the pentagon equation is also automatically satisfied.
A tensor category with a trivial associator is called a strict tensor category,
and there is a general theorem stating that any tensor category is equivalent to a strict tensor category.
Strict tensor categories are sometimes easier to work with,
since we can ignore the associator and the pentagon equation.
This is why endomorphism categories of algebras can be useful
in the study of tensor categories.

Given $t_1\in \Hom(\rho_1,\sigma_1)$ and $t_2\in \Hom(\rho_2,\sigma_2)$,
we also need to be able to form their tensor product 
$t_1\otimes t_2\in \Hom(\rho_1\otimes \rho_2,\sigma_1\otimes \sigma_2)$.
This is \emph{not} just a product in $\cA$, but is defined by \begin{equation}
    t_1\otimes t_2 := t_1 \rho_1(t_2) = \sigma_1(t_2) t_1.
\end{equation}
It is a good exercise to check that this indeed satisfies the intertwiner condition.

The discussions so far were formal, but other conditions 
require deeper structure theorems on $\cA$.
We want the unit object to be the identity endomorphism $\id:\cA\to \cA$.
For this to be simple, the center of $\cA$ should be trivial,
meaning that $\cA$ is a factor, in the terminology of the theory of operator algebras.
The existence of the direct sum $\rho\oplus\sigma$
and the dual $\rho^\vee$ requires a significantly more detailed analysis.
In the end, with suitable assumptions put on the infinite-dimensional algebra  $\cA$,
we can show that $\End(\cA)$ is a tensor category.
But the resulting category $\End(\cA)$ will contain
all of the endomorphisms of $\cA$ and will be huge.
To study a finite symmetry, or equivalently to consider a fusion category,
we typically consider a subcategory $\cC$ of $\End(\cA)$,
generated by a carefully chosen finite set of endomorphisms $\rho_X$.

\subsubsection{Relation to the construction of the `exotic' fusion categories}
Before proceeding, let us make some comments on the relevance 
of the above description to the construction of the `exotic' fusion categories.
The content of this subsection is tangential to the main topic of the paper,
so readers can skip it without affecting their understanding.

Let us first  recall what the `standard' fusion categories are.
As there are a number of known standard ways to modify a given fusion category to produce another, 
including but not limited to gauging (in the physics language),
which is the same as taking the bimodule category of an algebra object (in the mathematics language),
here we are going to classify fusion categories up to such modifications.

Then, there are two main sources of `standard' fusion categories,
namely 
\begin{itemize}
\item those coming from finite groups $G$, and
\item  those coming from the irreducible representations of
affine Lie algebras at some level, or equivalently from the Chern-Simons theory with a compact Lie group at some level.
\end{itemize}
For example, fusion categories whose simple objects are all invertible are called pointed fusion categories,
and they are classified by a finite group $G$ and a class in $H^3(BG;U(1))$,
as we will review in Sec.~\ref{sec:bosonic}.
The Ising category and the Fibonacci category are both obtained 
from the affine Lie algebras, the former from $\mathfrak{so}(9)$ level $1$,
and the latter from $\mathfrak{g}_2$ level $1$.

Until September 2026, fusion categories which belong to neither of these classes
were only finite in number (under a very rough classification
we are describing here), and such fusion categories can be called `exotic'.
The first example was the construction by Haagerup \cite{HaagerupOriginal} in the 1990s in the context of the study of subfactors,
whose details were presented in \cite{AsaedaHaagerup} together with another example.
A systematic way to try to generate more examples was given by Izumi \cite{Izumi1,Izumi2} in the 2000s, and a number of new examples were found.
This was again based on the study of endomorphism categories of certain
infinite-dimensional algebras.
Finally, in September 2026, a new infinite family of fusion categories
was found by three groups independently \cite{Huang:2026erd,RW2026,Gannon:2026enh}, 
again based on the method of Izumi.
For these reasons, the authors feel that the study of endomorphism categories of infinite-dimensional algebras is a very important tool in the study of fusion categories,
which is why the introductory material of this method was sketched here.
See also \cite{Yu:2025iqf} for another recent review of this material
aimed at the high-energy theory community.

\section{Bosonic case: pointed fusion categories}
\label{sec:bosonic}

We start with the bosonic case. Consider a pointed fusion category $\cC$,
i.e.~such that all simple objects are invertible.
We first show that the set $S(\cC)$ of isomorphism classes of simple objects in $\cC$ forms a group.
Take two simple objects $X$ and $Y$ in $\cC$.
Suppose $X\simeq X'$ and $Y\simeq Y'$.
Then $X\otimes Y\simeq X'\otimes Y'$.
This means that the tensor product $\otimes$ descends to $S(\cC)$.
This operation is associative, thanks to the existence of the associator
\begin{equation}
\alpha_{X,Y,Z}:(X\otimes Y)\otimes Z\xrightarrow{\sim} X\otimes (Y\otimes Z).
\end{equation}
The unit object $1$ provides the identity element of $S(\cC)$,
and the inverse of $X$ is given by $X^\vee$.
We now fix a finite group $G$ and consider a pointed fusion category $\cC$ such that $S(\cC)\simeq G$.
We would like to classify such $\cC$'s up to equivalence.\footnote{%
Here we consider the isomorphism $\varphi: S(\cC)\xrightarrow{\sim} G$ to be a part of the data,
so the equivalence of categories we consider is always assumed to respect this data.
Otherwise the classification results below all need to be considered up to the action of $\mathrm{Aut}\,G$.
}

\subsection{Cocycle conditions}

\paragraph{Introduction of a coordinate system:}
We first choose a representative $X_g$ for each $g\in G$.
Then, for each $g,h\in G$, we pick an isomorphism
\begin{equation}
u_{g,h}:X_g\otimes X_h\to X_{gh}.
\end{equation}
We call these choices a `coordinate system' for $\cC$.

\paragraph{Extraction of a 3-cochain:}
This allows us to consider the morphisms \begin{equation}
u_{g,hk}\circ (\id_{X_g}\otimes u_{h,k}): X_g\otimes (X_h\otimes X_k)\to X_{ghk}
\end{equation} and \begin{equation}
u_{gh,k}\circ (u_{g,h}\otimes \id_{X_k}): (X_g\otimes X_h)\otimes X_k\to X_{ghk}.
\end{equation}
We compare the two using the associator $\alpha_{X_g,X_h,X_k}$
\begin{equation}
\alpha_{X_g,X_h,X_k}: (X_g\otimes X_h)\otimes X_k\xrightarrow{\sim} X_g\otimes (X_h\otimes X_k).
\end{equation}
As the Hom space $\Hom(X_g\otimes (X_h\otimes X_k),X_{ghk})$ is one-dimensional, there should be a nonzero scalar $\omega(g,h,k)\in \bC^\times$ such that
\begin{equation}
u_{g,hk}\circ (\id_{X_g}\otimes u_{h,k})\circ \alpha_{X_g,X_h,X_k} = \omega(g,h,k) u_{gh,k}\circ (u_{g,h}\otimes \id_{X_k}).
\end{equation}
We abbreviate this equation by dropping $\id$ and $\circ$ and write it as \begin{equation}
u_{g,hk}\, u_{h,k}\, \alpha_{X_g,X_h,X_k} = \omega(g,h,k)\, u_{gh,k}\, u_{g,h},
\label{eq:omegadef}
\end{equation}
because $\id$ can be filled in unambiguously.

\paragraph{Cocycle condition on the 3-cochain:}
We now derive the cocycle condition on $\omega$. To do this,
we simply follow the pentagon relation
given in Fig.~\ref{fig:pentagon} very carefully.

We first consider 
the following manipulation:
\begin{align}
&u_{g,hkl} \, u_{h,kl}\, u_{k,l}\, \alpha_{X_h,X_k, X_l} \, \alpha_{X_g,X_h\otimes X_k,X_l}  \, \alpha_{X_g, X_h,X_k} \label{eq:start} \\
&=\omega(h,k,l)\, u_{g,hkl} \, u_{hk,l}\, u_{h,k}\, \alpha_{X_g,X_h\otimes X_k,X_l}  \, \alpha_{X_g,X_h,X_k}\\
&=\omega(h,k,l)\, u_{g,hkl} \, u_{hk,l} \,\alpha_{X_g,X_{hk},X_l}\,  u_{h,k} \,\alpha_{X_g,X_h,X_k}\\
&=\omega(h,k,l)\omega(g,hk,l) \,u_{ghk,l} \, u_{g,hk} \, u_{h,k}  \,\alpha_{X_g,X_h,X_k}\\
&=\omega(h,k,l)\omega(g,hk,l)\omega(g,h,k)\,u_{ghk,l} \,u_{gh,k}\, u_{g,h}.\label{eq:end0}
\end{align}
Next, we consider the following manipulation:
\begin{align}
&u_{g,hkl} \, u_{h,kl} \,u_{k,l}\,\alpha_{X_g,X_h,X_k\otimes X_l}\, \alpha_{X_g\otimes X_h,X_k,X_l}\label{eq:start2}\\
&=u_{g,hkl} \, u_{h,kl} \,\alpha_{X_g,X_h,X_{kl}}\, u_{k,l}\,\alpha_{X_g\otimes X_h,X_k,X_l}\\
&=\omega(g,h,kl)\,u_{gh,kl}\, u_{g,h}\, u_{k,l} \, \alpha_{X_g\otimes X_h,X_k,X_l}\label{eq:crucial1}\\
&=\omega(g,h,kl)\,u_{gh,kl}\, u_{k,l} \, u_{g,h}\,\alpha_{X_g\otimes X_h,X_k,X_l}\label{eq:crucial2}\\
&=\omega(g,h,kl)\,u_{gh,kl}\, u_{k,l} \, \alpha_{X_{gh},X_k,X_l}\, u_{g,h}\\
&=\omega(g,h,kl)\omega(gh,k,l)\,u_{ghk,l}\, u_{gh,k}\, u_{g,h}\label{eq:end}
\end{align}
Using the pentagon equation for the associator \begin{equation}
\alpha_{X_h,X_k, X_l} \, \alpha_{X_g,X_h\otimes X_k,X_l}  \, \alpha_{X_g, X_h,X_k}
= \alpha_{X_g,X_h,X_k\otimes X_l} \, \alpha_{X_g\otimes X_h,X_k,X_l},
\end{equation}
we see that 
the starting points of the two manipulations, Eqs.~\eqref{eq:start} and \eqref{eq:start2}, are the same.
Therefore, the two endpoints, Eqs.~\eqref{eq:end0} and \eqref{eq:end}, should also be the same, so
we find that
\begin{equation}
\omega(h,k,l)\omega(g,hk,l)\omega(g,h,k) = \omega(gh,k,l)\omega(g,h,kl).
\end{equation}
This indeed means that $\omega$ is a 3-cocycle in $Z^3(BG;U(1))$.

In general, any fusion category is equivalent to a skeletal one,
where there is a unique simple object in each isomorphism class,
or to a strict one, where the associator is trivial.
In the former case, there is no problem in declaring that $\Hom(X,X)=\bC$ 
for all simple objects $X$, and we can simply take all $u_{g,h}$ to be the identity.
In the latter case, all the associators $\alpha_{X,Y,Z}$ are the identity.
By making one of these choices, the manipulations we just performed become
significantly simplified.
Here, we chose not to do so, since the generalizations to the fermionic case are more straightforward in this form.

\paragraph{Dependence on the various choices involved:}
When we extracted the 3-cochain $\omega$,
we made various choices. Let us study how  these choices affect $\omega$.
There are three types of choices we made:
\begin{enumerate}
\item We could have chosen a different $\cC'$ equivalent to $\cC$.
\item We could have chosen a different representative $Y_g$ for each $g\in G$ in the same $\cC$.
\item We could have chosen different isomorphisms $v_{g,h}:X_g\otimes X_h\to X_{gh}$ in the same $\cC$ with the same representatives $X_g$.
\end{enumerate}
We investigate them in turn. 

On the first point, suppose we have a tensor functor $F:\cC\to \cC'$ giving an equivalence.
Suppose a coordinate system $X_g$ and $u_{g,h}$ is chosen for $\cC$.
Then, we can choose a coordinate system $X'_g=F(X_g)$ and $u'_{g,h}=F(u_{g,h})$ for $\cC'$.
Then, the 3-cochain $\omega'$ extracted from $\cC'$ is the same as $\omega$ extracted from $\cC$, i.e.~$\omega'=\omega$.

On the second point, suppose we have chosen a different representative $Y_g$ for each $g\in G$ in the same $\cC$.
Then, there is an isomorphism $f_g:X_g\to Y_g$ for each $g\in G$.
We can then choose $v_{g,h}:Y_g\otimes Y_h\to Y_{gh}$ as
\begin{equation}
v_{g,h}=f_{gh}\circ u_{g,h}\circ (f_g^{-1}\otimes f_h^{-1}).
\end{equation}
Then, the 3-cochain $\omega'$ extracted from $Y_g$ and $v_{g,h}$ turns out to be the same as $\omega$ extracted from $X_g$ and $u_{g,h}$, i.e.~$\omega'=\omega$.

On the third point, suppose we have chosen different isomorphisms $v_{g,h}:X_g\otimes X_h\to X_{gh}$ in the same $\cC$ with the same representatives $X_g$.
Then, there is a nonzero scalar $\lambda(g,h)\in \bC^\times$ such that
\begin{equation}
v_{g,h}=\lambda(g,h) u_{g,h}.\label{eq:lambda}
\end{equation}
Then, the 3-cochain $\omega'$ extracted from $X_g$ and $v_{g,h}$ is related to $\omega$ extracted from $X_g$ and $u_{g,h}$ as
\begin{equation}
\omega'(g,h,k)=\frac{\lambda(g,hk)\lambda(h,k)}{\lambda(g,h)\lambda(gh,k)}\omega(g,h,k).\label{eq:omegaprime}
\end{equation}
Written additively, this becomes \begin{equation}
    \omega'=\omega+\delta\lambda.
\end{equation}
This means that $\omega'$ and $\omega$ are in the same cohomology class in $H^3(BG;\bC^\times)\simeq H^3(BG;U(1))$.
Below, we will be cavalier 
and will not distinguish between $\bC^\times$ and $U(1)$
as coefficients of cohomology,
since the difference never affects us.

\paragraph{Uniqueness up to equivalence:}
Conversely, the class in $H^3(BG;U(1))$ determines the category up to equivalence.
Namely, suppose that two pointed fusion categories $\cC$ and $\cC'$ give the same class under chosen identifications $S(\cC)\simeq G\simeq S(\cC')$.
We show that they are equivalent.

Let $\omega$ and $\omega'$ be extracted from coordinate systems $(X_g,u_{g,h})$ of $\cC$ and $(X'_g,u'_{g,h})$ of $\cC'$.
As $\omega'$ and $\omega$ differ by a coboundary $\lambda$ by assumption, we can rescale the $u'_{g,h}$ as in \eqref{eq:lambda}
so that $\omega'=\omega$ exactly.
We now construct a tensor functor $F:\cC\to\cC'$.
We set $F(X_g)=X'_g$ and let $F$ act on $\Hom(X_g,X_g)=\bC$ as the identity;
as every object is a direct sum of the $X_g$'s, this determines $F$
as a linear functor.
A tensor functor also comes with natural isomorphisms $J_{X,Y}:F(X)\otimes F(Y)\to F(X\otimes Y)$,
and it suffices to specify them for simple objects.
We fix them by demanding $F(u_{g,h})\, J_{X_g,X_h}=u'_{g,h}$,
i.e.~$F$ maps the coordinate system of $\cC$ to that of $\cC'$.
We can straightforwardly check that $F$ is compatible with the associators, and hence is a tensor functor.
Therefore $\cC\simeq\cC'$.

\subsection{Construction}

So far we showed that we can extract a class in $H^3(BG;U(1))$ from a pointed fusion category $\cC$ with $S(\cC)\simeq G$.
Let us do the converse.
Namely,  given a finite group $G$ and a 3-cocycle $\omega'\in Z^3(BG;U(1))$, we construct a skeletal pointed fusion category $\cC(G,\omega')$ as follows.
We declare that there is a unique simple object $X_g$ for each $g\in G$,
$\Hom(X_g,X_g)$ is a copy of $\bC$,
the tensor product is $X_g\otimes X_h=X_{gh}$,
and the associator is given by $\alpha_{X_g,X_h,X_k}=\omega'(g,h,k)\id_{X_{ghk}}$.
The pentagon identity for $\alpha$ is equivalent to the cocycle condition for $\omega'$.
Taking $u_{g,h}=\id_{X_{gh}}$ and extracting $\omega(g,h,k)$ as above, we find that $\omega(g,h,k)=\omega'(g,h,k)$.
Therefore every class in $H^3(BG;U(1))$ arises from some pointed fusion category.

Combined with the uniqueness shown above,
any pointed fusion category $\cC$ with $S(\cC)\simeq G$ is equivalent to $\cC(G,\omega)$,
where $\omega$ is extracted from $\cC$.
We conclude that such categories, up to equivalence, are classified by $H^3(BG;U(1))$.
For example, when $G=\bZ_2$, we have $H^3(BG;U(1))=\bZ_2$.

Before moving on, let us note that this construction gives a skeletal fusion category.
In Appendix~\ref{app:spinchain}, we construct instead a strict fusion category,
using the technique of half-space symmetry actions on the algebra of local operators;
the spin chain is the edge of a group-cohomology model,
whose symmetry action was described in general terms by Else and Nayak \cite[Appendix~E]{Else:2014vma}.

\subsection{Addition law}

Given two pointed fusion categories $\cC$ and $\cD$ with $S(\cC)\simeq S(\cD)\simeq G$, 
we construct a new pointed fusion category $\cC \square \cD$ with $S(\cC\square \cD)\simeq G$ as follows.
We first take the Deligne tensor product $\cC\boxtimes \cD$.\footnote{
    For physicists, this is a category-level operation
    analogous to the product of two groups,
    constructing $G\times G'$ from $G$ and $G'$.
    For example, the simple objects of $\cC\boxtimes \cD$ are of the form $X\boxtimes Y$ where $X\in \cC$ and $Y\in \cD$ are simple objects.
    The following construction is an operation analogous to 
    taking the diagonal subgroup $G\subset G\times G$.
}
Then, we take the full subcategory of $\cC\boxtimes \cD$ 
generated by the simple objects $X\boxtimes Y$ where $X\in \cC$ and $Y\in \cD$
are such that they correspond to the same group element $g\in G$.
We denote this full subcategory by $\cC\square \cD$.

Let us now suppose that coordinate systems $(X_g,u_{g,h})$ of $\cC$ and $(Y_g,v_{g,h})$ of $\cD$ are chosen.
Then we can choose a coordinate system of $\cC\square \cD$ as $(X_g\boxtimes Y_g,u_{g,h}\boxtimes v_{g,h})$.
Let $\omega_\cC(g,h,k)$ and $\omega_\cD(g,h,k)$ be the 3-cochains extracted from these coordinate systems.
Then, the 3-cochain $\omega_{\cC\square \cD}(g,h,k)$ extracted from this coordinate system is easily seen to be given by
\begin{equation}
\omega_{\cC\square \cD}(g,h,k)=\omega_\cC(g,h,k)\omega_\cD(g,h,k).
\end{equation}
Therefore, the operation $\square$ on pointed fusion categories corresponds to the addition law in $H^3(BG;U(1))$.
The unit is $\omega=1$, realized by $\cC(G,1)$,
and the inverse of $\omega$ is $\bar\omega(g,h,k)=\omega(g,h,k)^{-1}$, realized by $\cC(G,\bar\omega)$.

\section{Fermionic case I: pointed super fusion categories}
\label{sec:fermionicI}

Let us move on to the fermionic case. We consider super fusion categories rather than fusion categories.
The difference is that the Hom space $\Hom(X,Y)$ between two objects $X$ and $Y$ is a super vector space, i.e.~it has a $\bZ_2$-grading: \begin{equation}
\Hom(X,Y)=\Hom(X,Y)_0\oplus \Hom(X,Y)_1.
\end{equation}
Physically, the even part $\Hom(X,Y)_0$ and the odd part $\Hom(X,Y)_1$ correspond to the space of bosonic operators and fermionic operators, respectively.
In the following, we will only consider morphisms of definite degree,
and we will denote the degree of a morphism $f$ by $|f|\in \bZ_2=\{0,1\}$.
We impose the condition that the associator $\alpha_{X,Y,Z}$ is always even, 
i.e.~$|\alpha_{X,Y,Z}|=0$.

For convenience, we introduce a simple object $\Pi$ such that
$\Hom(1,\Pi)$ and $\Hom(\Pi,1)$ are one-dimensional and odd, and $\Pi\otimes \Pi\simeq 1$.
It is known that we can always formally adjoin such an object $\Pi$ to any super fusion category $\cC$ if it does not have one already \cite{BrundanEllis},
and we assume its existence in the following.

We now consider a pointed super fusion category $\cC$, i.e.~such that all simple objects are invertible.
Then $\Hom(X,X)$ for a simple object $X$ is $\bC$,
and $\Hom(X,Y)$ for two simple objects $X$ and $Y$ is zero- or one-dimensional.
We can then repeat the same analysis as in Sec.~\ref{sec:bosonic},
with only a minor modification.

The content of this part of the paper has also appeared in many physics papers;
see e.g.~\cite{Cheng:2015yoa,Seifnashri:2023dpa,Okada:2025kie,Chavda:2026eeq}.

\subsection{Cocycle conditions}
\paragraph{Introduction of a coordinate system:}
As before, the set $S(\cC)$ of isomorphism classes of simple objects in $\cC$ forms a group.
We fix a finite group $G$ and fix the isomorphism $S(\cC)\simeq G$.
Then,
we choose a representative $X_g$ for each $g\in G$ and an isomorphism $u_{g,h}:X_g\otimes X_h\to X_{gh}$ for each $g,h\in G$.
They form a coordinate system for $\cC$.

\paragraph{Extraction of cochains:}
We first define a $\bZ_2$-valued 2-cochain $\nu(g,h)$ by the relation
\begin{equation}
    \nu(g,h)=|u_{g,h}|.
\end{equation}
We also define a 3-cochain $\omega(g,h,k)\in \bC^\times$ from 
the relation \eqref{eq:omegadef} as before.

\paragraph{Conditions on the cochains:}
The 2-cochain $\nu$ is a cocycle, that is, it satisfies the condition
\begin{equation}
\nu(g,hk)+\nu(h,k)=\nu(gh,k)+\nu(g,h).
\end{equation}
This follows by taking the degree of both sides of \eqref{eq:omegadef} and using the fact that $\alpha_{X_g,X_h,X_k}$ is even.

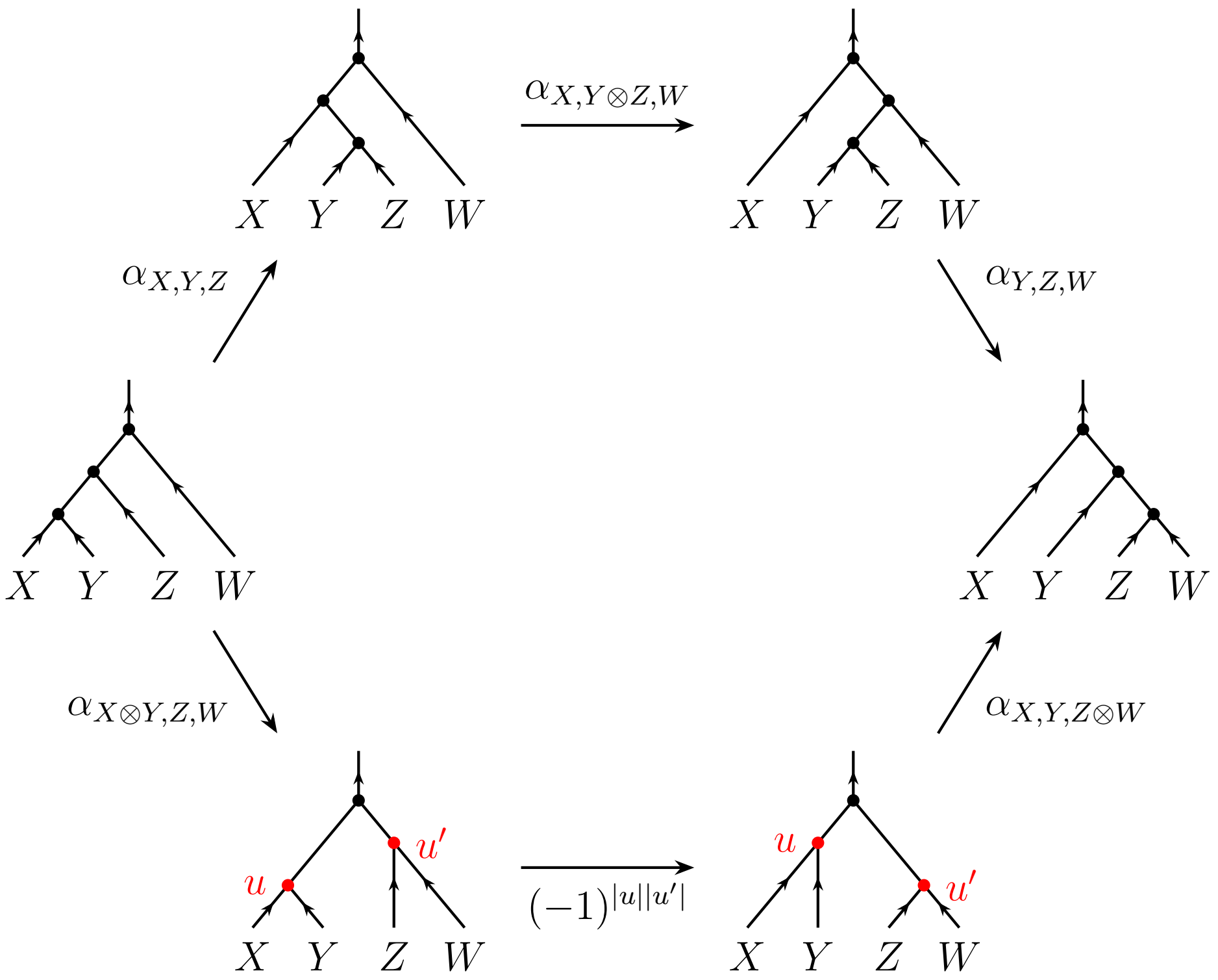
\begin{figure}
    \centering
    \begin{tikzpicture}[scale=0.9]
        \begin{scope}[shift={(-5.4,0)},local bounding box=A]
            \fourtreecoords
            \draw[sline] (L1) -- (N12); \draw[sline] (L2) -- (N12);
            \draw[thick] (N12) -- (N13); \draw[sline] (L3) -- (N13);
            \draw[thick] (N13) -- (R); \draw[sline] (L4) -- (R);
            \treeblobs{N12,N13,R}
        \end{scope}
        \begin{scope}[shift={(-2.8,4.2)},local bounding box=B]
            \fourtreecoords
            \draw[sline] (L2) -- (N23); \draw[sline] (L3) -- (N23);
            \draw[sline] (L1) -- (N13); \draw[thick] (N23) -- (N13);
            \draw[thick] (N13) -- (R); \draw[sline] (L4) -- (R);
            \treeblobs{N23,N13,R}
        \end{scope}
        \begin{scope}[shift={(2.8,4.2)},local bounding box=C]
            \fourtreecoords
            \draw[sline] (L2) -- (N23); \draw[sline] (L3) -- (N23);
            \draw[thick] (N23) -- (N24); \draw[sline] (L4) -- (N24);
            \draw[sline] (L1) -- (R); \draw[thick] (N24) -- (R);
            \treeblobs{N23,N24,R}
        \end{scope}
        \begin{scope}[shift={(-2.8,-4.2)},local bounding box=D1]
            \fourtreecoords
            \coordinate (N34) at (0.4,-0.04); 
            \draw[sline] (L1) -- (N12); \draw[sline] (L2) -- (N12);
            \draw[sline] (L3) -- (N34); \draw[sline] (L4) -- (N34);
            \draw[thick] (N12) -- (R); \draw[thick] (N34) -- (R);
            \treeblobs{R}
            \node[blob,red,label={[red]left:$u$}] at (N12) {};
            \node[blob,red,label={[red]right:$u'$}] at (N34) {};
        \end{scope}
        \begin{scope}[shift={(2.8,-4.2)},local bounding box=D2]
            \fourtreecoords
            \coordinate (N12) at (-0.4,-0.04); 
            \draw[sline] (L1) -- (N12); \draw[sline] (L2) -- (N12);
            \draw[sline] (L3) -- (N34); \draw[sline] (L4) -- (N34);
            \draw[thick] (N12) -- (R); \draw[thick] (N34) -- (R);
            \treeblobs{R}
            \node[blob,red,label={[red]left:$u$}] at (N12) {};
            \node[blob,red,label={[red]right:$u'$}] at (N34) {};
        \end{scope}
        \begin{scope}[shift={(5.4,0)},local bounding box=E]
            \fourtreecoords
            \draw[sline] (L3) -- (N34); \draw[sline] (L4) -- (N34);
            \draw[sline] (L2) -- (N24); \draw[thick] (N34) -- (N24);
            \draw[sline] (L1) -- (R); \draw[thick] (N24) -- (R);
            \treeblobs{N34,N24,R}
        \end{scope}
        \begin{scope}[-{Stealth[length=2mm]},thick,shorten >=6pt,shorten <=6pt]
            \draw (A) -- node[above left] {$\alpha_{X,Y,Z}$} (B);
            \draw (B) -- node[above] {$\alpha_{X,Y\otimes Z,W}$} (C);
            \draw (C) -- node[above right] {$\alpha_{Y,Z,W}$} (E);
            \draw (A) -- node[below left] {$\alpha_{X\otimes Y,Z,W}$} (D1);
            \draw (D1) -- node[below] {$(-1)^{|u||u'|}$} (D2);
            \draw (D2) -- node[below right] {$\alpha_{X,Y,Z\otimes W}$} (E);
        \end{scope}
    \end{tikzpicture}
    \caption{The pentagon relation of Fig.~\ref{fig:pentagon} in the fermionic case.
    In the lower path,
    we need to reverse the order of the multiplication of the two junctions
    labeled by $u$ (joining $X$ and $Y$) and $u'$ (joining $Z$ and $W$),
    shown in red in the picture.
    This exchange produces the sign $(-1)^{|u||u'|}$.}
    \label{fig:fermionic-sign}
\end{figure}

The condition on $\omega$ can be found by repeating the manipulations \eqref{eq:start}--\eqref{eq:end} we performed in Sec.~\ref{sec:bosonic}.
The only difference is that there is an additional sign factor
$(-1)^{\nu(g,h)\nu(k,l)}$ in the manipulation from \eqref{eq:crucial1} to \eqref{eq:crucial2},
which comes from the fact that two fermionic operators on different lines anticommute with each other, see Fig.~\ref{fig:fermionic-sign}.
The final result is that $\omega$ and $\nu$ satisfy the following condition:
\begin{equation}
\omega(h,k,l)\omega(g,hk,l)\omega(g,h,k)=(-1)^{\nu(g,h)\nu(k,l)}\omega(gh,k,l)\omega(g,h,kl).
\end{equation}
Written additively, by regarding $U(1)\simeq \bR/\bZ$, this condition reads
\begin{equation}
    \delta \omega = \frac12 \nu \cup \nu,\label{eq:GW}
\end{equation}
where $\frac12$ is the natural map $\bZ_2\to \bR/\bZ$ sending $1\mapsto 1/2$.

\paragraph{Dependence on the various choices involved:}
As in the bosonic case, there are various choices involved in the extraction of $\nu$ and $\omega$.
As before, the choices of equivalent super fusion categories do not affect $\nu$ and $\omega$.
When we choose different representatives $Y_g$ for each $g\in G$ in the same $\cC$, 
the degree of $f_g:X_g\to Y_g$ affects both $\nu$ and $\omega$.
Let us write the new 2-cochain and 3-cochain as $\nu'$ and $\omega'$, respectively.
We also use the 1-cochain $\xi(g):=|f_g|$.
Then we first find 
\begin{equation}
\nu'(g,h)=\nu(g,h)+\xi(g)+\xi(h)-\xi(gh),
\end{equation} that is, $\nu'$ and $\nu$ are in the same cohomology class in $H^2(BG;\bZ_2)$.
We also find that
\begin{equation}
\omega'(g,h,k)=(-1)^{\xi(g)\nu'(h,k)+\nu(g,h)\xi(k)}\omega(g,h,k).
\end{equation}

Finally, when we choose different isomorphisms $v_{g,h}:X_g\otimes X_h\to X_{gh}$ in the same $\cC$ with the same representatives $X_g$ as in \eqref{eq:lambda},
we find that $\nu$ is unchanged while $\omega$ is shifted by a coboundary,
so that 
we have \begin{equation}
\omega'=\omega+\delta \lambda,
\end{equation}
as in \eqref{eq:omegaprime}.

We use the notation $SH^3(BG)$ for the set of pairs $(\nu,\omega)$ satisfying \eqref{eq:GW} modulo the two equivalences discussed above.
This is the supercohomology group of Gu and Wen, first introduced in \cite{Gu:2012ib},
although we have not discussed the group structure of $SH^3(BG)$ yet.

\paragraph{Uniqueness up to equivalence:}
The argument is the same as in the bosonic case, with two additional points.
Suppose $(\nu,\omega)$ and $(\nu',\omega')$, extracted from $\cC$ and $\cC'$, define the same element of $SH^3(BG)$.
First, every equivalence of the cochains is realized by an actual change of the coordinate system of $\cC'$.
Indeed,  given a 1-cochain $\xi$,
we can take the new representatives $Y_g=\Pi^{\xi(g)}\otimes X'_g$,
with $f_g:X'_g\to Y_g$ an odd isomorphism when $\xi(g)=1$.
Also, given a $\lambda$, we can rescale the $u'_{g,h}$.
We can therefore assume $\nu'=\nu$ and $\omega'=\omega$ exactly.

Second,
we define $F:\cC\to\cC'$ by $F(X_g)=X'_g$  and $F(\Pi)=\Pi$.
The natural isomorphisms are defined
by $F(u_{g,h})\,J_{X_g,X_h}=u'_{g,h}$ as before.
As $F(u_{g,h})$ and $u'_{g,h}$ have degrees $\nu(g,h)$ and $\nu'(g,h)$,
$J_{X_g,X_h}$ is even because $\nu=\nu'$.
Then, the compatibility of $F$ with the associators can be checked as before,
using the fact that $\omega=\omega'$.
(We again omit the discussion of the objects $\Pi\otimes X_g$.)
Therefore $\cC\simeq\cC'$.

\subsection{Construction}
Let us now start from a pair of cochains and construct a pointed super fusion category.
Given a finite group $G$ and a pair of cochains $(\nu,\omega)$ satisfying \eqref{eq:GW}, we construct a pointed super fusion category $\cC(G,\nu,\omega)$ as follows.
We declare that there are two simple objects $X_g$ and $\Pi\otimes X_g=X_g\otimes\Pi$ for each $g\in G$,
$\Hom(X_g,X_g)$ is a copy of $\bC$,
$\Hom(X_g,\Pi\otimes X_g)$ is a copy of $\bC$ in the odd part which we denote by $\bC \psi$ with $\psi$ being an odd basis element.
$\Hom(X_g,X_h)$ is zero-dimensional for $g\neq h$.
We declare that the composition of morphisms is given by the usual multiplication in $\bC$ and the standard multiplication rules of 
the odd element $\psi$.

We define the tensor product as follows:
\begin{equation}
X_g\otimes X_h = \begin{cases}
X_{gh} & \text{if $\nu(g,h)=0$},\\
\Pi\otimes X_{gh} & \text{if $\nu(g,h)=1$}.
\end{cases}
\end{equation}
This assigns isomorphic objects to $(X_g\otimes X_h)\otimes X_k$ and $X_g\otimes (X_h\otimes X_k)$ 
thanks to the cocycle condition on $\nu$.
This rule also makes sure that $\Hom(X_g\otimes X_h,X_{gh})$ 
is one-dimensional and
is purely even or purely odd depending on the value of $\nu(g,h)$.
We then choose $u_{g,h}$ to be $1$ or $\psi$ depending on the value of $\nu(g,h)$.
Finally, we define the associator $\alpha_{X_g,X_h,X_k}$ 
in terms of $\omega(g,h,k)$ as in \eqref{eq:omegadef}.

Strictly speaking, we need to specify $\alpha_{X_g,X_h,\Pi X_k}$,
$\alpha_{X_g,\Pi X_h,X_k}$, etc., as well.
It is known that this can be done \cite{BrundanEllis,Usher}, and we omit the details here.

With the coordinate system $(X_g,u_{g,h})$ chosen above,
the cochains extracted from $\cC(G,\nu,\omega)$ are the original $(\nu,\omega)$.
Combined with the uniqueness shown above,
any pointed super fusion category $\cC$ with $S(\cC)\simeq G$
is equivalent to $\cC(G,\nu,\omega)$, where $(\nu,\omega)$ is extracted from $\cC$.
We conclude that such categories, up to equivalence, are classified by $SH^3(BG)$.

\subsection{Addition law}

Suppose now that we have two pointed super fusion categories $\cC$ and $\cD$ with $S(\cC)\simeq S(\cD)\simeq G$,
and we have extracted $(\nu_\cC,\omega_\cC)$ and $(\nu_\cD,\omega_\cD)$ from them,
by choosing coordinate systems $(X_g,u_{g,h})$ and $(Y_g,v_{g,h})$.
We can construct a new pointed super fusion category $\cC\square \cD$ as before,
from the coordinate system $(X_g\boxtimes Y_g,u_{g,h}\boxtimes v_{g,h})$.
Then, the 2-cochain and 3-cochain can be extracted from this coordinate system. 
As $|u_{g,h}\boxtimes v_{g,h}|=|u_{g,h}|+|v_{g,h}|$, we find that
\begin{equation}
\nu_{\cC\square \cD}(g,h)=\nu_\cC(g,h)+\nu_\cD(g,h).
\end{equation}
Next, we compare \begin{multline}
(u_{g,hk}\boxtimes v_{g,hk})(u_{h,k}\boxtimes v_{h,k})\alpha_{X_g\boxtimes Y_g,X_h\boxtimes Y_h,X_k\boxtimes Y_k}\\
=\omega_{\cC\square \cD}(g,h,k)(u_{gh,k}\boxtimes v_{gh,k})(u_{g,h}\boxtimes v_{g,h})
\end{multline}
against \begin{multline}
    (u_{g,hk}u_{h,k}\alpha_{X_g,X_h,X_k})\boxtimes
    (v_{g,hk}v_{h,k}\alpha_{Y_g,Y_h,Y_k})\\
    =\omega_\cC(g,h,k)\omega_\cD(g,h,k)
    (u_{gh,k}u_{g,h})\boxtimes (v_{gh,k}v_{g,h}),
\end{multline}
using the interchange law
\begin{equation}
(a\boxtimes b)(a'\boxtimes b')=(-1)^{|b||a'|}(aa'\boxtimes bb').
\end{equation}
We  find that
\begin{equation}
\omega_{\cC\square \cD}(g,h,k)=\omega_\cC(g,h,k)\omega_\cD(g,h,k)
(-1)^{\nu_\cC(h,k)\nu_\cD(g,hk)+\nu_\cC(g,h)\nu_\cD(gh,k)},
\end{equation}
which, written additively, is \begin{equation}
    \nu_{\cC\square \cD}=\nu_\cC+\nu_\cD,\qquad
    \omega_{\cC\square \cD}=\omega_\cC+\omega_\cD+\frac12\nu_\cD\cup_1\nu_\cC,
    \label{eq:SH3-addition}
\end{equation}
where $\cup_1$ is the cup-1 product of Steenrod, given for 2-cochains by
\begin{equation}
(x\cup_1 y)(g,h,k)=x(g,hk)y(h,k)+x(gh,k)y(g,h).
\label{eq:cup1}
\end{equation}
This is the standard addition law in the supercohomology group $SH^3(BG)$,
which we can also write as \begin{equation}
(\nu,\omega)\square(\nu',\omega')=(\nu+\nu',\omega+\omega'+\frac12\nu'\cup_1\nu).
\end{equation}
Note that at the cochain level,
$(\nu,\omega)\square(\nu',\omega')$
and $(\nu',\omega')\square(\nu,\omega)$ differ 
by $\frac12(\nu\cup_1\nu'-\nu'\cup_1\nu)=\frac12\delta(\nu\cup_2\nu')$,
which is a coboundary.
Therefore the group law is abelian at the cohomology level.

We note that the abelian group $SH^3(BG)$ has two homomorphisms  \begin{equation}
H^3(BG;U(1))\to SH^3(BG)\to H^2(BG;\bZ_2),
\end{equation}  the first of which sends $\omega$ to $(0,\omega)$ and 
the second of which sends $(\nu,\omega)$ to $\nu$.
This is not  a short exact sequence in general, however.
Although the first map is known to be injective,
the second is not necessarily surjective;
it only hits classes $\nu$ such that the image of $\nu\cup\nu$ under $H^4(BG;\bZ_2)\to H^4(BG;U(1))$ vanishes.
That said, an explicit computation shows that
for $G=\bZ_2$ it is a short exact sequence, and $SH^3(B\bZ_2)=\bZ_4$;
note that $\nu\cup\nu\neq0$ in $H^4(B\bZ_2;\bZ_2)$ for the nontrivial $\nu$, but $H^4(B\bZ_2;U(1))=0$.

\section{Fermionic case II: weakly-pointed super fusion categories}
\label{sec:fermionicII}

So far we have assumed that every simple object is invertible.
However, this assumption excludes the possibility of having Majorana zero modes on the topological defect lines.
Indeed, if a simple object $X$ has a Majorana zero mode, so that
$\End(X)=\Cl(\psi)=\bC\oplus \bC\psi$,
its dual $X^\vee$ also has a Majorana zero mode, so that
$\End(X^\vee)=\Cl(\psi^\vee)=\bC\oplus \bC\psi^\vee$.
Then, the tensor product $X\otimes X^\vee$ has two Majorana zero modes,
with an action of $\Cl(\psi,\psi^\vee)$,
which has two projection operators $P_\pm=\frac12(1\pm i\psi\psi^\vee)$.
This means that $X\otimes X^\vee$ can always be decomposed into 
two subobjects, say $Z$ and $Z'$.
Furthermore, $\psi$ and $\psi^\vee$ provide odd isomorphisms between $Z$ and $Z'$, 
because $\psi P_+ =P_-\psi$, and so on.

The minimal generalization is then to demand that both these subobjects 
are isomorphic to the unit object $1$, but the two are related by an odd isomorphism. This means that
\begin{equation}
X\otimes X^\vee\simeq 1\oplus \Pi,
\end{equation}
where we remind the reader that $\Pi\simeq 1$  is a simple object isomorphic to $1$
but with an odd isomorphism $\psi:1\to \Pi$.
In this case, $X$ is ever so slightly non-invertible,
to the minimal extent required by the existence of a Majorana zero mode.
We call a simple object $X$ satisfying this condition \emph{weakly invertible},
with the understanding that invertible objects are also weakly invertible.
We then call a super fusion category \emph{weakly-pointed} if all its simple objects are weakly invertible.

The aim of the rest of the paper is to 
classify weakly-pointed super fusion categories, 
and to show that the result is compatible with the known classification of fermionic anomalies in 1+1 dimensions using spin bordism groups.
In this section we perform the classification,
and in the next section we compare the result with the spin bordism group.

\subsection{Preparations}
Now, let us consider a weakly-pointed super fusion category $\cC$,
and let $S(\cC)$ be the set of isomorphism classes of simple objects in $\cC$.
This naturally forms a group, as in the pointed case.
The only nontrivial case is when we have two simple objects $X$ and $Y$ with Majorana zero modes, so that $X\otimes Y\simeq Z\oplus Z'$.
Denoting $\End(X)=\Cl(\psi)$, $\End(Y)=\Cl(\psi')$, 
$Z$ and $Z'$ are projections by $P_\pm=\frac12(1\pm i\psi\psi')$.
Then, $\psi$ and $\psi'$ provide odd isomorphisms between $Z$ and $Z'$, 
so that $Z\simeq Z'$.
Then, the group structure on $S(\cC)$ is given by $[X]\cdot [Y]=[Z]$.
It is straightforward to check that this operation indeed defines a group.

Now, we fix a finite group $G$ and an isomorphism $S(\cC)\simeq G$.
We would like to classify such $\cC$'s up to equivalence.
We need to introduce a coordinate system,
extract cochains, and study them.
The main complication is that
the pentagon equation is now a matrix equation
on $\End(X_g\otimes X_h\otimes X_k\otimes X_l)$, 
which is no longer one-dimensional;
when all of $X_g$, $X_h$, $X_k$, $X_l$ have Majorana zero modes,
it is a Clifford algebra with four generators, which is 16-dimensional.
We need to find a way to extract a scalar equation from this matrix equation.\footnote{%
In particular, the reference \cite{Usher}, 
although generally quite instructive,
does not seem to provide a proper treatment of this issue.}
There are multiple ways to do this,
and the system can be analyzed in any of them,
but we found the following approach to be the most convenient.

The essential idea is the following.
Consider the tensor product of a number of simple objects $X_{g_1}\otimes\cdots\otimes X_{g_n}$.
This will have a number of Majorana zero modes, say $k$.
In the analysis of the pentagon equation and so on,
we need to control multiple morphisms in a single Hom space
\begin{equation}
V:=\Hom(X_{g_1}\otimes\cdots\otimes X_{g_n},Y),
\end{equation}
where $Y$ is (a direct sum of a number of copies of)
a simple object $X_{g_1\cdots g_n}$.
It is convenient to concentrate on the case
when $Y$ is isomorphic to $X_{g_1}\otimes\cdots\otimes X_{g_n}$.
In this case, the Hom space $V$ has an action
of the Clifford algebra $A:=\Cl(\psi_1,\ldots,\psi_k)$ from the right,
and also another action of $A':=\End(Y)$ from the left,
and $A\simeq A'$,
both by composition of the morphisms.
That is, $V$ is an $(A',A)$-bimodule.
Now, by counting the dimensions, we find that,
as a super vector space, $V\simeq A\simeq A'$.

This means that the situation is simply modeled by 
the Clifford algebra $A$ with $k$ generators,
acted on from the left and the right by itself,
\emph{once an explicit isomorphism $A\simeq A'$ is chosen}.
Once this is done, the space of elements $v\in V$
which graded-commute with the left and the right actions of $A$,
that is, which satisfy \begin{equation}
a v =(-1)^{|a||v|} v a,\qquad a\in A,
\end{equation}
is one-dimensional and spanned by the identity element $1\in V\simeq A$.
So far in our analysis, a `coordinate system' consisted of $X_g$ for each $g$
and $u_{g,h}$ for each $g,h$.
In the weakly-pointed case, 
we need to add as the data the explicit isomorphisms of $\End(Y)$ appearing 
in various steps to a fixed Clifford algebra.
This allows us to put all the morphisms we consider
 in this one-dimensional subspace.
Then we can extract a scalar equation from the matrix equation.

\subsection{Cocycle conditions}

\paragraph{The coordinate system and extraction of cochains:}
As before, we choose a representative $X_g$ for each $g\in G$.
We let $m(g)\in\{0,1\}$ be the number of Majorana zero modes on $X_g$,
regarded as an integer, rather than modulo 2.
Since the Majorana zero modes of $X_g\otimes X_h$ pair up when $m(g)=m(h)=1$,
$m$ modulo 2 defines a homomorphism $G\to\bZ_2$, i.e.~$\delta m=0$ modulo 2.
When $m(g)=1$, we have $\End(X_g)\simeq \Cl(\psi_g)$.
Here, we \emph{choose and fix} an explicit isomorphism,
and regard $\psi_g\in \End(X_g)$ as a specific odd operator with $\psi_g^2=1$.

We also take an object $M\simeq 1\oplus \Pi$
such that $\End(M)\simeq \Cl(\chi,\chi')$,
and we pick an explicit isomorphism, so that $\chi$ and $\chi'$ are specific anticommuting odd operators squaring to $1$.
For definiteness we demand that $\frac12(1-i\chi\chi')$ is the projector onto the summand $1$ of $M$.
As $M$ is built out of $1$ and $\Pi$ only, it can be moved past any object $Y$ by
the canonical isomorphism $c_{M,Y}:M\otimes Y\to Y\otimes M$
coming from the canonical isomorphisms for $1$ and $\Pi$,
$1\otimes Y=Y=Y\otimes 1$ and $\Pi\otimes Y\simeq Y\simeq Y\otimes \Pi$.\footnote{%
We define this latter morphism to be $(\id_Y \otimes \psi)(\psi^{-1} \otimes \id_Y)$,
which is independent of the nonzero odd element $\psi\in \Hom(1,\Pi)$.
With this definition $c_{M,Y}$ is natural also for odd morphisms.
\label{foot:cMY}
}
We will use these canonical identifications without further comment;
we also suppress the associators involving $M$.

We now define an auxiliary object \begin{equation}
X_{g,h}:=X_{gh}\otimes M^{\otimes m(g)m(h)},
\end{equation}
so that $X_g\otimes X_h\simeq X_{g,h}$ for all $g,h\in G$.
Then $\End(X_g\otimes X_h)$ and $\End(X_{g,h})$ are both
Clifford algebras with $m(g)+m(h)$ generators,
and we identify them explicitly as follows:
when $m(g)=m(h)=1$, we identify $\psi_g$ and $\psi_h$ with $\chi$ and $\chi'$ acting on $M$, respectively;
when only one of $m(g)$, $m(h)$ is $1$, we identify the corresponding $\psi_g$ or $\psi_h$ with $\psi_{gh}$.
We  choose an (even or odd) isomorphism \begin{equation}
u_{g,h}:X_g\otimes X_h\to X_{g,h}
\end{equation} so that $u_{g,h}$
graded-commutes
with the Majorana operators on both sides under this identification,
e.g.~$u_{g,h}\,\psi_g=(-1)^{|u_{g,h}|}\chi\, u_{g,h}$ when $m(g)=m(h)=1$.
The space of such isomorphisms is one-dimensional, as we discussed above,
so the choice of $u_{g,h}$ is unique up to a nonzero scalar.
We let $\nu(g,h)\in \bZ_2$ be the degree of $u_{g,h}$, as before.

Now, we generalize the definition of $X_{g,h}$ above
to triple products, by setting
\begin{equation}
X_{g,(h,k)}:=X_{g,hk}\otimes M^{\otimes m(h)m(k)},\qquad
X_{(g,h),k}:=X_{gh,k}\otimes M^{\otimes m(g)m(h)}.
\end{equation}
We then have the isomorphisms
\begin{align}
u_{g,hk}\,u_{h,k}&:X_g\otimes(X_h\otimes X_k)\to X_g\otimes X_{hk}\otimes M^{\otimes m(h)m(k)}\to X_{g,(h,k)},\\
u_{gh,k}\,u_{g,h}&:(X_g\otimes X_h)\otimes X_k\to X_{gh}\otimes M^{\otimes m(g)m(h)}\otimes X_k\to X_{(g,h),k},
\end{align}
where in the second line we moved $M^{\otimes m(g)m(h)}$ to the right of $X_k$ before applying $u_{gh,k}$.
Both $X_{g,(h,k)}$ and $X_{(g,h),k}$ are $X_{ghk}$
possibly tensored with  $M$,
and their endomorphism algebras are identified with the Clifford algebra generated by $\psi_g$, $\psi_h$, $\psi_k$
through the identifications above.
These two identifications literally agree, except when $m(g)=m(h)=m(k)=1$.

In this last case, both objects are $X_{ghk}\otimes M$,
and the triple $(\psi_g,\psi_h,\psi_k)$ is identified with $(\psi_{ghk},\chi,\chi')$ on $X_{g,(h,k)}$
but with $(\chi,\chi',\psi_{ghk})$ on $X_{(g,h),k}$.
We denote by $\iota_{g,h,k}:X_{g,(h,k)}\to X_{(g,h),k}$ the isomorphism 
which reorders the Clifford generators involved.
Namely, $\iota_{g,h,k}$ is the identity unless $m(g)=m(h)=m(k)=1$,
in which case it is the even operator on $X_{ghk}\otimes M$ satisfying
\begin{equation}
\iota\,\psi_{ghk}=\chi\,\iota,\quad 
\iota\,\chi=\chi'\,\iota,\quad 
\iota\,\chi'=\psi_{ghk}\,\iota.
\end{equation}
Such an operator is unique up to a scalar $\epsilon$:
\begin{equation}
\iota_{g,h,k}=\epsilon\cdot \frac12\bigl(1-\psi_{ghk}\chi-\chi\chi'+\psi_{ghk}\chi'\bigr)
\qquad\text{when $m(g)=m(h)=m(k)=1$.}
\label{eq:iota}
\end{equation}

We can finally define a 3-cochain $\omega(g,h,k)\in \bC^\times$ from the relation
\begin{equation}
\iota_{g,h,k}\, u_{g,hk}\, u_{h,k}\, \alpha_{X_g,X_h,X_k} = \omega(g,h,k)\, u_{gh,k}\, u_{g,h}.
\label{eq:omegadef-weak}
\end{equation}
This is possible, because
both sides take values in a one-dimensional subspace of $\Hom(X_g\otimes X_h\otimes X_k,X_{(g,h),k})$ of definite parity
graded-commuting with the Majorana operators $\psi_g$, $\psi_h$, $\psi_k$ on both sides.

\paragraph{Remarks on the choice of $\iota$:}
Fixing $\iota_{g,h,k}$ precisely requires various choices:
the object $M$ together with its Majorana operators $\chi,\chi'$,
the convention on which summand of $M$ is $1$,
the order in which the factors of $M$ are placed in $X_{g,(h,k)}$ and $X_{(g,h),k}$,
the use of the canonical isomorphisms $c_{M,Y}$,
and also the overall normalization $\epsilon$ in \eqref{eq:iota}.
The operators $\iota$ appear in the pentagon equation discussed below,
and a different choice can easily introduce additional signs or phases
in the condition satisfied by $\omega$.
We will show later in Sec.~\ref{sec:fermion-II-construction}
that the choice
\begin{equation}
\epsilon=e^{2\pi i/8}
\label{eq:epsilon}
\end{equation}
together with the conventions stated above is a particularly good one,
which makes the condition on $\omega$ as simple as possible.

\paragraph{Conditions on the cochains:}
Let us actually derive the conditions on the cochains.
We already remarked that $m$ is a homomorphism $G\to\bZ_2$, i.e.~$\delta m=0$.
As for $\nu$, taking the degree of both sides of \eqref{eq:omegadef-weak} and using the fact that $\iota_{g,h,k}$ and the associator are even,
we find that $\nu$ is a cocycle as before:
\begin{equation}
\nu(g,hk)+\nu(h,k)=\nu(gh,k)+\nu(g,h),
\quad\text{that is,}\quad \delta\nu=0.
\end{equation}

To find the condition on $\omega$, we repeat the manipulations \eqref{eq:start}--\eqref{eq:end0}
and \eqref{eq:start2}--\eqref{eq:end},
with the objects $X_{g,(h,(k,l))}$, $X_{((g,h),k),l}$ etc.~defined recursively in the same manner.
There are two differences from the pointed case.
First, every time we use the defining relation \eqref{eq:omegadef-weak} of $\omega$,
an operator $\iota$ appears:
$\iota_{h,k,l}$, $\iota_{g,hk,l}$ and $\iota_{g,h,k}$ in the first manipulation,
and $\iota_{g,h,kl}$ and $\iota_{gh,k,l}$ in the second.
As these $\iota$'s are even polynomials of the Majorana operators,
and the $u$'s graded-commute with the Majorana operators as we assumed,
we can move all the $\iota$'s to the left end without generating any signs.
Second, the copies of $M$ produced by the $u$'s need to be moved to the right,
which also does not generate any signs.\footnote{%
This uses the naturality of $c_{M,Y}$ also for odd morphisms mentioned in footnote~\ref{foot:cMY};
in particular, no sign depending on $\nu$ arises here.
}
The sign $(-1)^{\nu(g,h)\nu(k,l)}$ from the step \eqref{eq:crucial1}--\eqref{eq:crucial2} is there as before.
Therefore, we end up with the relation
\begin{equation}
\omega(h,k,l)\omega(g,hk,l)\omega(g,h,k)\, I_1=(-1)^{\nu(g,h)\nu(k,l)}\omega(gh,k,l)\omega(g,h,kl)\, I_2,
\end{equation}
where $I_1$ is the composition of $\iota_{h,k,l}$, $\iota_{g,hk,l}$, $\iota_{g,h,k}$
and $I_2$ is the composition of $\iota_{g,h,kl}$ and $\iota_{gh,k,l}$,
appropriately tensored with identities and transported to $X_{g,(h,(k,l))}$.

Both $I_1$ and $I_2$ are isomorphisms $X_{g,(h,(k,l))}\to X_{((g,h),k),l}$
implementing the same rearrangement of the Majorana operators,
and therefore \begin{equation}
I_1=\kappa(m(g),m(h),m(k),m(l)) I_2
\end{equation}
for a nonzero scalar $\kappa(m(g),m(h),m(k),m(l))$,
implying that $\omega$ satisfies the condition
\begin{multline}
\kappa(m(g),m(h),m(k),m(l))\omega(h,k,l)\omega(g,hk,l)\omega(g,h,k)\\
=(-1)^{\nu(g,h)\nu(k,l)}\omega(gh,k,l)\omega(g,h,kl).
\label{eq:kappa}
\end{multline}

Note that the scalars $\kappa$ appearing here are independent 
of the specific category $\cC$ under consideration
and depend only on our conventions on the choice of $\iota$
and how we insert and move the copies of $M$ around.
With the choice \eqref{eq:epsilon}, it turns out that we 
have $\kappa(m(g),m(h),m(k),m(l))=1$ for all choices of $m(g)$, $m(h)$, $m(k)$, $m(l)$, as we will see later.
Assuming this, $\omega$ satisfies the same condition as in the pointed case,
i.e.~we have
\begin{equation}
\delta m=0,\qquad \delta\nu=0,\qquad \delta \omega = \frac12 \nu \cup \nu
\label{eq:weakGW}
\end{equation}
in the additive notation.

\paragraph{Dependence on the various choices involved:}
First, any two choices of $M$ with its two Majorana operators
within $\cC$ are related by an even isomorphism compatible with the Majorana operators,
which is unique up to a scalar,
and the $\iota$'s, being made from $M$, are transported along it.
Therefore, we are not going to discuss the dependence of the cochains on these choices.

Then the choices we would like to discuss are:
the choice of a super fusion category among equivalent ones,
and then the choice of a `coordinate system',
that is, the choice of the representatives $X_g$ for $g\in G$,
the Majorana operators $\psi_g$ on $X_g$ when $m(g)=1$,
and the isomorphisms $u_{g,h}$.
Let us go over each one of them.

Equivalent super fusion categories give the same cochains.
Rescaling $u_{g,h}$ by $\lambda(g,h)$ changes $\omega$ by $\delta\lambda$ as in \eqref{eq:omegaprime}.

The most complicated case is the change of the representatives $X_g$,
together with the Majorana operators $\psi_g$.
This is done by isomorphisms $f_g:X_g\to Y_g$ of degree $\xi(g)$.
Then $\psi_g^Y$ is determined by the condition
$f_g\,\psi_g^X=(-1)^{\xi(g)}\psi_g^Y f_g$
from $\psi_g^X$ and $f_g$.
Note that flipping the sign of $\psi_g$ is a special case, obtained by taking $f_g=\psi_g$ itself,
which is odd.
As $\iota_{g,h,k}$ in \eqref{eq:iota} is expressed in terms of the Majorana operators,
it is automatically transported to the corresponding operator for $Y_g$'s,
and the computation goes through as in the pointed case, giving
\begin{align}
\nu'(g,h)&=\nu(g,h)+\xi(g)+\xi(h)-\xi(gh),\\
\omega'(g,h,k)&=(-1)^{\xi(g)\nu'(h,k)+\nu(g,h)\xi(k)}\omega(g,h,k).
\end{align}

To summarize, a weakly-pointed super fusion category $\cC$ with $S(\cC)\simeq G$ determines
a triple $(m,\nu,\omega)$ satisfying \eqref{eq:weakGW},
modulo the equivalences generated by $\xi$ and $\lambda$.
When $m=0$, this reduces to the pointed case $SH^3(BG)$ discussed in Sec.~\ref{sec:fermionicI}.

\paragraph{Uniqueness up to equivalence:}
As in the pointed cases, the triple determines the category up to equivalence.
Suppose that the triples $(m,\nu,\omega)$ and $(m',\nu',\omega')$ extracted from $\cC$ and $\cC'$
are related by the equivalences above.
The argument is the same as in Sec.~\ref{sec:fermionicI},
with a few additional points concerning the Majorana operators.

First, we can assume that the triples agree exactly.
Indeed, $m=m'$, since $m(g)$ is the number of Majorana zero modes on the simple object labeled by $g$.
The equivalences generated by $\xi$ and $\lambda$ are realized by actual changes of the coordinate system of $\cC'$ as before,
where the Majorana operators $\psi'_g$ are transported along the isomorphisms $f_g$.

Second, we define $F:\cC\to\cC'$ by $F(X_g)=X'_g$ and $F(\Pi)=\Pi$.
When $m(g)=1$, $\End(X_g)=\Cl(\psi_g)$ is not just $\bC$,
and we specify the action of $F$ on it by $F(\psi_g)=\psi'_g$.
As $M$ is built out of $1$ and $\Pi$ only, $F$ maps $M$ to $M$, together with its Majorana operators $\chi$, $\chi'$
and the canonical isomorphisms $c_{M,Y}$.
The natural isomorphisms $J_{X_g,X_h}$ are fixed by $F(u_{g,h})\,J_{X_g,X_h}=u'_{g,h}$ as before,
where we identify $F(X_{g,h})$ with $X'_{g,h}$ in the obvious manner.
They are even because $\nu=\nu'$.
A new point is that $J_{X_g,X_h}$ also needs to be natural with respect to the Majorana operators,
i.e.~it should intertwine $\psi'_g\otimes 1$ and $1\otimes\psi'_h$ with $F(\psi_g\otimes 1)$ and $F(1\otimes\psi_h)$.
This holds, since both $u_{g,h}$ and $u'_{g,h}$ graded-commute with the Majorana operators
under the same identification of $\psi_g$, $\psi_h$ with $\chi$, $\chi'$ or $\psi_{gh}$,
and $F$ maps each of these operators to its counterpart in $\cC'$.

Finally, the compatibility of $F$ with the associators is checked as before,
now using \eqref{eq:omegadef-weak} in $\cC$ and in $\cC'$.
The only new ingredient is $\iota_{g,h,k}$,
but it is given by the same expression \eqref{eq:iota} in terms of the Majorana operators in both categories,
and therefore $F$ maps $\iota_{g,h,k}$ in $\cC$ to $\iota_{g,h,k}$ in $\cC'$.
Then the compatibility condition reduces to $\omega=\omega'$, as in the pointed case.
(We again omit the discussion of the objects $\Pi\otimes X_g$.)
Therefore $\cC\simeq\cC'$.

Note that this argument does not use the condition \eqref{eq:weakGW},
nor the specific choice \eqref{eq:epsilon} of the normalization of $\iota$;
it only uses the fact that $\omega$ is defined by the same relation \eqref{eq:omegadef-weak} in both categories.
In particular, it does not require the reference category $\cR$ introduced below.

\subsection{Construction}
\label{sec:fermion-II-construction}
Next we would like to show that we can construct a weakly-pointed super fusion category from any triple $(m,\nu,\omega)$ satisfying \eqref{eq:weakGW}.
For this we need to introduce the reference category $\cR$.

\paragraph{The reference category:}
The reference category $\cR$ is a weakly-pointed super fusion category with $S(\cR)=\bZ_2$,
whose only simple object not isomorphic to $1$ is
a weakly invertible object $\sigma$ with $\End(\sigma)=\Cl(e)$.
We construct it in Appendix~\ref{app:reference}, where it is called $\cR_1$; here we only need the following properties:
\begin{itemize}
\item There are objects $\sigma^n$ for $n\ge 0$ such that
$\sigma^a\otimes \sigma^b=\sigma^{a+b}$ strictly, with trivial associators.
\item 
$\End(\sigma^n)=\Cl(e_1,\ldots,e_n)$.
\item For $x\in\End(\sigma^a)$ and $y\in\End(\sigma^b)$, 
$x\otimes y\in\End(\sigma^{a+b})$ is determined by
$e_k\otimes 1=e_k$ and $1\otimes e_k=e_{a+k}$.
\item The object $M:=\sigma^2$ is isomorphic to $1\oplus\Pi$,
with $\chi=e_1$ and $\chi'=e_2$ satisfying our convention that $\frac12(1-i\chi\chi')$ projects onto $1$.
\item 
Finally, the canonical isomorphism $c_{M,\sigma}:M\otimes \sigma\to \sigma\otimes M$, which is an endomorphism of $\sigma^3$, is given by
\begin{equation}
c_{M,\sigma}=e^{2\pi i/8}\cdot\frac12\bigl(1-e_1e_2-e_2e_3+e_1e_3\bigr).
\label{eq:cMsigma}
\end{equation}
\end{itemize}

Let us first apply our extraction procedure to $\cR$ itself
to get a triple $(m,\nu,\omega)$ of cochains.
We use the additive notation for the group $\bZ_2=\{0,1\}$,
and set  $X_0=1$, $X_1=\sigma$, $\psi_1=e_1$ and $M=\sigma^2$.
Then $X_1\otimes X_1=\sigma^2=X_{0}\otimes M$,
with $e_1$, $e_2$ identified with $\chi$, $\chi'$ as required.
We can take all $u_{g,h}$ to be the identity, so that $\nu=0$.
For $g=h=k=1$, both $X_{g,(h,k)}$ and $X_{(g,h),k}$ are $\sigma\otimes M=\sigma^3$,
with $(\psi_{ghk},\chi,\chi')=(e_1,e_2,e_3)$.
Here $u_{g,hk}u_{h,k}$ is the identity,
while $u_{gh,k}u_{g,h}$ is $c_{M,\sigma}$, since we need to move $M$ past $X_k=\sigma$.
Then the defining relation \eqref{eq:omegadef-weak} becomes $\iota_{1,1,1}=\omega(1,1,1)c_{M,\sigma}$.
Comparing \eqref{eq:iota} and \eqref{eq:cMsigma},
we see that the choice \eqref{eq:epsilon} gives $\omega(1,1,1)=1$.
In all the other cases $\omega=1$ trivially.
Therefore the reference category $\cR$ has $(\nu,\omega)=(0,1)$.

We can use this to show that the proportionality factor $\kappa(m(g),m(h),m(k),m(l))$ in \eqref{eq:kappa} is always $1$.
Indeed, as already argued, this factor is independent of the specific category $\cC$, so we can compute it in $\cR$.
For $\cR$, all possible values of $(m(g),m(h),m(k),m(l))$ occur,
and $\nu=0$, $\omega=1$.
Therefore $\kappa=1$ always, as claimed above.

\paragraph{Construction from cochains:}
Now, given a finite group $G$ and a triple $(m,\nu,\omega)$ satisfying \eqref{eq:weakGW},
we construct a weakly-pointed super fusion category $\cC(G,m,\nu,\omega)$ as follows.
We first take the pointed super fusion category $\cC(G,\nu,\omega)$ constructed in Sec.~\ref{sec:fermionicI},
whose simple objects we now denote by $Z_g$, with isomorphisms $u^Z_{g,h}:Z_g\otimes Z_h\to Z_{gh}$ of degree $\nu(g,h)$.
We then consider the Deligne tensor product $\cC(G,\nu,\omega)\boxtimes\cR$,
and define $\cC(G,m,\nu,\omega)$ to be its full subcategory generated by
\begin{equation}
X_g:=Z_g\boxtimes \sigma^{m(g)}
\end{equation}
under direct sums and direct summands.
Clearly $\End(X_g)=\Cl(\psi_g)$ with $\psi_g=1\boxtimes e_1$ when $m(g)=1$ and $\End(X_g)=\bC$ otherwise.
Furthermore, as $m(gh)+2m(g)m(h)=m(g)+m(h)$, we have
\begin{equation}
X_g\otimes X_h=(Z_g\otimes Z_h)\boxtimes\sigma^{m(g)+m(h)}
\simeq 
X_{gh}\otimes M^{\otimes m(g)m(h)},
\end{equation}
where we now regard $M=1\boxtimes\sigma^2$ as an object of $\cC(G,m,\nu,\omega)$.
In particular every simple object is weakly invertible, and $S(\cC(G,m,\nu,\omega))=G$.
As a coordinate system we take $X_g$, $\psi_g$, $M$ as above and
$u_{g,h}=u^Z_{g,h}\boxtimes\id_{\sigma^{m(g)+m(h)}}$,
which graded-commutes with the Majorana operators thanks to the interchange law, and has degree $\nu(g,h)$.
Then all the operators acting on the $\cR$ part
(the identities, $c_{M,\sigma}$, and $\iota$) are even,
and the computation of \eqref{eq:omegadef-weak} factorizes into the one in $\cC(G,\nu,\omega)$,
which gives $\omega(g,h,k)$, and the one in $\cR$, which gives $1$ as we just saw.
Therefore the cochains extracted from $\cC(G,m,\nu,\omega)$ are the original $(m,\nu,\omega)$.

Combined with the uniqueness shown above,
any weakly-pointed super fusion category $\cC$ with $S(\cC)\simeq G$
is equivalent to $\cC(G,m,\nu,\omega)$, where $(m,\nu,\omega)$ is extracted from $\cC$.
For a rigorous treatment, see \cite{Kawasumi}.

\subsection{Addition law}
Given two weakly-pointed super fusion categories $\cC$ and $\cC'$ with $S(\cC)\simeq S(\cC')\simeq G$,
we define $\cC\square\cC'$ as before, as the full subcategory of $\cC\boxtimes\cC'$
generated by the objects $X_g\boxtimes X'_g$ and their direct summands.
When $m(g)=m'(g)=1$, the object $X_g\boxtimes X'_g$ has two Majorana operators $\psi_g\boxtimes 1$ and $1\boxtimes \psi'_g$,
and splits into two simple objects which are oddly isomorphic to each other;
we take one of them as the representative.
Therefore $\cC\square\cC'$ is again weakly-pointed with $S(\cC\square\cC')=G$,
and its $m$ is $m+m'$ modulo 2.

The cochains of $\cC\square\cC'$ are extracted from the coordinate system induced from those of $\cC$ and $\cC'$.
By the uniqueness shown above, we can perform the computation assuming $\cC=\cC(G,m,\nu,\omega)$ and $\cC'=\cC(G,m',\nu',\omega')$,
since the equivalences constructed there map the coordinate systems to each other.
Then, using the symmetry of the Deligne product, which only introduces the standard Koszul signs, we have
\begin{equation}
X_g\boxtimes X'_g=(Z_g\boxtimes \sigma^{m(g)})\boxtimes(Z'_g\boxtimes\sigma^{m'(g)})
\simeq (Z_g\boxtimes Z'_g)\boxtimes(\sigma^{m(g)}\boxtimes\sigma^{m'(g)}).
\end{equation}
The first factor generates the pointed super fusion category $\cC(G,\nu,\omega)\square\cC(G,\nu',\omega')$ studied in Sec.~\ref{sec:fermionicI}.
The second factor lives in the universal category 
$\cR\boxtimes\cR$.
This is a weakly-pointed super fusion category with group $\bZ_2\times\bZ_2$ and $m(a,b)=a+b$ modulo 2,
whose simple objects are $Y_{0,0}=1\boxtimes 1$,
$Y_{1,0}=\sigma\boxtimes 1$, $Y_{0,1}=1\boxtimes\sigma$, and 
$Y_{1,1}$ one of the simple summands of $\sigma\boxtimes\sigma$.
Let us denote its cochains by $(\nu_{\cR\cR},\omega_{\cR\cR})$.

Then the cochains of $\cC\square \cC'$ are obtained by combining three pieces:
those of $\cC(G,\nu,\omega)$, those of $\cC(G,\nu',\omega')$, and the pull-back of
$(\nu_{\cR\cR},\omega_{\cR\cR})$ by the map $\mathsf{m}:=(m,m'):G\to\bZ_2\times\bZ_2$.
The rule to combine them is the same as in the pointed case \eqref{eq:SH3-addition},
since the derivation there only used the interchange law,
and the additional operators $\iota$ and $c_{M,Y}$ we have now are all even.
Therefore the addition law for the cochains of $\cC\square\cC'$ is
\begin{equation}
(m,(\nu,\omega))\square(m',(\nu',\omega'))=
(m+m',(\nu,\omega)\square(\nu',\omega')\square\mathsf{m}^*(\nu_{\cR\cR},\omega_{\cR\cR})).
\label{eq:sum-formula}
\end{equation}

We determine $(\nu_{\cR\cR},\omega_{\cR\cR})$ in Appendix~\ref{app:RR},
where it is shown that we can take
\begin{equation}
\nu_{\cR\cR}(g,h)=x(g)y(h),\qquad
\omega_{\cR\cR}(g,h,k)=i^{x(g)y(h)y(k)}(-1)^{x(g)x(h)y(h)y(k)},
\label{eq:nuomegaRR}
\end{equation}
where $x(a,b)=a$ and $y(a,b)=b$ for $(a,b)\in\bZ_2\times\bZ_2$.
Written additively, this gives \begin{equation}
\mathsf{m}^*(\nu_{\cR\cR},\omega_{\cR\cR})
=\Bigl(m\cup m',\ \frac12\, m\cup(m\cup_1 m')\cup m'+\frac14\,\tilde m\cup\tilde m'\cup\tilde m'\Bigr),
\label{eq:sum-formula2}
\end{equation}
where $(m\cup_1 m')(g)=m(g)m'(g)$,
and $\tilde m$, $\tilde m'$ are the lifts of $m$, $m'$ to integral cochains taking values in $\{0,1\}$.

To summarize, weakly-pointed super fusion categories with $S(\cC)\simeq G$ are classified by
triples $(m,\nu,\omega)$ satisfying \eqref{eq:weakGW} modulo the equivalences described above,
and the operation $\square$ makes them into an abelian group,
as $\square$ is commutative and associative up to equivalence, by the corresponding properties of the Deligne product.
At the level of cochains, the addition law is given by
\eqref{eq:sum-formula} with $\mathsf{m}^*(\nu_{\cR\cR},\omega_{\cR\cR})$ given by \eqref{eq:sum-formula2}.
Denote the resulting abelian group by $SSH^3(BG)$,
although this is not a standard notation.
Our results show that this is a group extension of the form \begin{equation}
0\longrightarrow SH^3(BG)\longrightarrow SSH^3(BG)\longrightarrow H^1(BG;\bZ_2)\longrightarrow 0,
\end{equation}
where the first map is the inclusion with $m=0$
and the second map is the projection to $m$;
the extension is determined by $(\nu_{\cR\cR},\omega_{\cR\cR})$.
This group is $\bZ_8$ for $G=\bZ_2$.

We will compare this with the result of Brumfiel and Morgan in the next section.
Before doing so, we note that our formula depends on various choices
we made in the construction.
For example, if we perform the replacement $i\mapsto -i$ in the construction
of the reference category $\cR$ in Appendix~\ref{app:reference},
the last factor in \eqref{eq:sum-formula2} would have the coefficient
$-1/4$ instead of $1/4$.

\section{Comparison with the spin bordism group}
\label{sec:comparison}

\subsection{Summary of results on the geometric side}
Now that we have classified weakly-pointed super fusion categories,
we would like to compare the result with the known classification of fermionic anomalies in 1+1 dimensions using spin bordism groups.
The modern general theory of anomalies,
given e.g.~in \cite{Freed:2016rqq,Yonekura:2018ufj},
states that the anomalies of a $(d+1)$-dimensional fermionic quantum field theory with a finite symmetry group $G$ are classified by
the Pontryagin dual of the spin bordism group $\tilde\Omega_{d+2}^{\text{spin}}(BG)$.
Such Pontryagin duals were explicitly determined in terms of
cochain data on $BG$ for $d+2\le 4$
in \cite{BrumfielMorgan1,BrumfielMorgan2}.
What we need is the case $d+2=3$; the result is summarized as follows.
We use the notation of \cite{BrumfielMorgan1},
except that we write $\delta$ for the coboundary and $\cup$ for the cup product explicitly.

For $X=BG$ we use group cochains, for which their $\cup_1$ product becomes \eqref{eq:cup1}.
We consider a triple of cochains $(w,p,a)$ on $BG$,
a 3-cochain $w\in C^3(BG;\bR/\bZ)$, a 2-cochain $p\in C^2(BG;\bZ_2)$, and a 1-cochain $a\in C^1(BG;\bZ_2)$,
where $w$ is considered only modulo coboundaries. 
They impose the following cocycle conditions:
\begin{equation}
\delta a=0,\qquad \delta p=0,\qquad \delta w=\frac12 p\cup p.
\label{eq:BM-cocycle}
\end{equation}
The product of two triples is
\begin{equation}
(w,p,a)\cdot(v,q,b)=(u,\ p+q+a\cup b,\ a+b),
\label{eq:BM-addition}
\end{equation}
where \begin{equation}
u=w+v+\frac12\Bigl[p\cup_1q+(p+q)\cup_1(a\cup b)+a\cup(a\cup_1 b)\cup b\Bigr]+\frac14 A\cup B\cup B.
\end{equation}
Here $A,B\in C^1(BG;\bZ)$ are the lifts of $a,b$ to integral cochains taking values in $\{0,1\}$,
$\frac14$ is the map $\bZ\to\bR/\bZ$ sending $1\mapsto 1/4$,
and $(a\cup_1 b)(g)=a(g)b(g)$.
This operation preserves the cocycle conditions \eqref{eq:BM-cocycle}.

The equivalences are generated by multiplication by triples of the form
\begin{equation}
\mathbf{D}'(t,x)=\Bigl(\frac12 t\cup\delta t,\ \delta t,\ \delta x\Bigr),
\end{equation}
where $t\in C^1(BG;\bZ_2)$ and $x\in C^0(BG;\bZ_2)$.
For group cochains, $C^0(BG;\bZ_2)=\bZ_2$ and $\delta x=0$, so $x$ plays no role here,
and the equivalence reads
\begin{equation}
(w,p,a)\sim (w+\frac12 t\cup\delta t+\frac12p\cup_1\delta t,\ p+\delta t,\ a).
\label{eq:BM-equiv}
\end{equation}
The product is not commutative at the level of cochains, but it becomes commutative and associative on equivalence classes.
The main theorem of \cite{BrumfielMorgan1} is that the resulting abelian group of equivalence classes is isomorphic to $\Hom(\tilde\Omega_3^{\text{spin}}(BG),\bR/\bZ)$.

\subsection{Comparison with the categorical result}
Let us now compare this with what we found in Sec.~\ref{sec:fermionicII}.
We map our data $(m,\nu,\omega)$ to the triple $(w,p,a)$ of \cite{BrumfielMorgan1} by
\begin{equation}
a=m,\qquad p=\nu,\qquad w=\omega,
\end{equation}
where $\omega$ is written additively, i.e.~regarded as valued in $\bR/\bZ\simeq U(1)$.
Our $\xi$ corresponds to their $t$,
and the gauge transformations by $\lambda$ correspond to the fact that $w$ is considered only modulo coboundaries.
Then the cocycle conditions \eqref{eq:weakGW} and \eqref{eq:BM-cocycle} are the same.
Using $\delta(p\cup_1t)=p\cup t+t\cup p+p\cup_1\delta t$,
we see that the shift of $w$ in \eqref{eq:BM-equiv}
agrees up to a coboundary with $\frac12(t\cup p'+p\cup t)$ where $p'=p+\delta t$,
which is what we found in Sec.~\ref{sec:fermionicI} and Sec.~\ref{sec:fermionicII}.

Our addition law found at the end of Sec.~\ref{sec:fermionicII} reads, in this notation,
\begin{equation}
u=w+v+\frac12\Bigl[q\cup_1p+(a\cup b)\cup_1(p+q)+a\cup(a\cup_1 b)\cup b\Bigr]+\frac14 A\cup B\cup B.
\end{equation}
The terms involving only $a$ and $b$ agree on the nose with \eqref{eq:BM-addition},
while the two terms of the form $\frac12 x\cup_1 y$ appear with the order of $x$ and $y$ exchanged.
As $x\cup_1 y+y\cup_1 x=\delta(x\cup_2 y)$ modulo 2 for cocycles $x$, $y$,
the difference is a coboundary,
which is immaterial since $w$ is considered only modulo coboundaries.

We note that the sign of the last term $\frac14 A\cup B\cup B$ 
was  fixed in \cite{BrumfielMorgan1} by an explicit computation of the Arf invariant on $\bR P^3$,
which depends on the orientation and spin structure conventions.
As discussed at the end of Sec.~\ref{sec:fermionicII}, the sign of our last term $\frac14 A\cup B\cup B$ also depends on our convention $P=\frac12(1-ie_1e_2)$,
and the complex conjugate convention would give $-\frac14 A\cup B\cup B$.
Thus, matching the sign in front of $1/4$ with \cite{BrumfielMorgan1} 
depends heavily on conventions on both sides.
Comparing this sign intrinsically would require computing a spin bordism invariant, such as the value on $\bR P^3$,
directly from the categorical data, which we do not attempt here,
although it would indeed be an interesting problem
to directly construct
spin bordism invariants from weakly-pointed super fusion categories,
and vice versa.

With this caveat understood, the classification of weakly-pointed super fusion categories with $S(\cC)\simeq G$
perfectly agrees with $\Hom(\tilde\Omega_3^{\text{spin}}(BG),\bR/\bZ)$, including the group structure.
This is the main result of this paper.

\section*{Acknowledgments}

KK is supported by FoPM, WINGS Program, the University of Tokyo and JST SPRING, Japan Grant Number JPMJSP2108.
YT is supported in part by WPI Initiative, MEXT, Japan at Kavli IPMU, the University of Tokyo
and by JSPS KAKENHI Grant-in-Aid (Kiban-C), No.24K06883.

LLMs were used to fill in the details of some calculations,
constructions and proofs,
but the overall structure of the paper is due to the authors.
The authors checked all the output of the LLMs;
they also edited the LLM output heavily so that 
it is as readable as what the authors would have written
before the advent of researcher-level LLMs.

\appendix

\section{Spin chain realization of pointed fusion categories}
\label{app:spinchain}

In this appendix,
we consider a concrete spin chain with an action of a finite group $G$ specified by a 3-cocycle $\alpha\in Z^3(BG;U(1))$,\footnote{%
In this appendix we reuse $\alpha$ for the 3-cocycle,
as the category we construct here is strict and 
no associator is needed, which frees up the symbol $\alpha$ for us.}
study the algebra of local operators,
and show that its automorphisms realize the pointed fusion category $\cC(G,\alpha)$.
This is to illustrate the general discussion of Sec.~\ref{sec:symmetry-algebra}
 in the context of Sec.~\ref{sec:bosonic}.

The particular spin chain model we consider here
is a  generalization to arbitrary $G$ and $\alpha$
of a model for $G=\bZ_2$ and $\alpha$ nontrivial given in \cite{Chen:2011pg},
whose general form in arbitrary dimensions was also given in \cite[Appendix~E]{Else:2014vma}.
The same model can also be obtained by applying the `anyon chain' or `golden chain' construction of \cite{Feiguin:2006ydp} 
to the pointed fusion category $\cC(G,\alpha)$.
It is also to be pointed out that a quite similar construction 
of automorphisms on a von Neumann algebra of type II$_1$
for general $G$ and $\alpha$ was given in classic papers on operator algebras
\cite{Jones:1979,Jones:1980},
generalizing the construction of Connes \cite{Connes:1977} for $G=\bZ_n$.

\paragraph{The spin chain and the algebra:}
Consider an infinite spin chain with sites $i\in \bZ$,
where each site has the local Hilbert space $\bC^{|G|}$ spanned by $|a\rangle$, $a\in G$.
As we only consider the right half of the chain,
from now on we restrict our attention to the sites $i\ge1$,
and write a basis state of the chain as $|a_1,a_2,\ldots\rangle$.
Let $\cA=\bigotimes_{i\ge1}M_{|G|}(\bC)$ be the algebra of operators acting on finitely many sites, 
i.e.~for any element $x\in\cA$, there exists $N$ such that $x$ acts as the identity on all sites $i>N$.

We write $\delta^{(i)}_a=|a\rangle\langle a|$ for convenience.
Then, for a function $F$ on $G$, we write $F(\hat a_i)=\sum_a F(a)\delta^{(i)}_a$
for the corresponding diagonal operator at the $i$-th site.

\paragraph{The automorphisms:}
Let $\alpha\in Z^3(BG;U(1))$.
For $g\in G$, we formally define
\begin{equation}
U_g|a_1,a_2,\ldots\rangle=\prod_{i\ge1}\alpha(g,a_{i+1},a_{i+1}^{-1}a_i)\;|ga_1,ga_2,\ldots\rangle.
\label{eq:EN-action}
\end{equation}
This infinite product is difficult to deal with rigorously.
But each factor involves only two neighboring sites,
and therefore \begin{equation}
\rho_g(x)=U_gxU_g^{-1}
\end{equation}
defines a well-defined automorphism of $\cA$.
To make this argument more precise, we can restrict the product
in the definition of $U_g$ to $i=1,\ldots,N$,
where $N$ is large enough so that $x$ acts as the identity on all sites $i>N$.
Then the product is finite, and $\rho_g(x)$ is independent of $N$.
We do not repeat this argument below.

\paragraph{The intertwiners:}
Let us compute $U_gU_h$.
The phase associated with the bond $(i,i+1)$ in $U_gU_h$ is
$\alpha(h,k,l)\alpha(g,hk,l)$ with $k=a_{i+1}$ and $l=a_{i+1}^{-1}a_i$,
since $(ha_{i+1})^{-1}ha_i=a_{i+1}^{-1}a_i$.
By the cocycle condition, this equals $\alpha(gh,k,l)$ times $\alpha(g,h,a_i)/\alpha(g,h,a_{i+1})$.
Multiplying over all $i\ge 1$, the factors $\alpha(g,h,a_{i+1})$ cancel except for the leftmost one, and we have
\begin{equation}
U_gU_h\propto U_{gh}\,\alpha(g,h,\hat a_1).
\end{equation}
We therefore have \begin{equation}
\rho_{gh}(x)=u_{g,h}\,\rho_g(\rho_h(x))\,u_{g,h}^{-1},\qquad
u_{g,h}=\alpha(g,h,(gh)^{-1}\hat a_1)^{-1}\in\Hom(\rho_g\circ\rho_h,\rho_{gh}).
\end{equation}
This element $u_{g,h}$ is a local unitary of $\cA$ 
at the left end of the chain, as was illustrated in Fig.~\ref{fig:half-line-action}.

\paragraph{The category:}

Let $\cC$ be the subcategory of $\End(\cA)$ generated by the $\rho_g$
(where we have to formally add the direct sums in this case).
The center of $\cA$ is trivial:
an element $x\in\cA$ acts on finitely many sites, say $1,\ldots,N$,
and if it commutes with everything in $\cA$, it commutes in particular with the full matrix algebra $M_{|G|}(\bC)^{\otimes N}=M_{|G|^N}(\bC)$ acting on these sites,
so it is a scalar.
Therefore $\Hom(\rho_e,\rho_e)=Z(\cA)=\bC$.

An argument similar to the above shows that $\Hom(\rho_g,\rho_g)$
is only the scalars.
When $g\neq h$,  $x\in\Hom(\rho_g,\rho_h)$ satisfies $x\rho_g(y)=\rho_h(y)x$ for all $y\in\cA$.
But for any $x$, there is an $N$ such that $x$ acts as the identity on all sites $i>N$, but $\rho_g$ and $\rho_h$ act differently on $\cA$
for arbitrarily large $i$, so $x$ must be zero.
So we find that $\rho_g$ for $g\in G$ are distinct simple objects,
and therefore $\cC$ is a pointed fusion category with $S(\cC)\simeq G$.

\paragraph{Extraction of the 3-cocycle:}

Let us use $u_{g,h}$ as the coordinate system of Sec.~\ref{sec:bosonic}
and extract the 3-cocycle $\omega$ as described in that section.
As $\End(\cA)$ is strict, the defining relation \eqref{eq:omegadef} becomes
\begin{equation}
u_{g,hk}\,\rho_g(u_{h,k})=\omega(g,h,k)\,u_{gh,k}\,u_{g,h}.
\end{equation}
Using
$
\rho_g(F(\hat a_1))=F(g^{-1}\hat a_1)
$
and setting $z=(ghk)^{-1}\hat a_1$,
this becomes \begin{equation}
\alpha(g,hk,z)^{-1}\alpha(h,k,z)^{-1}=\omega(g,h,k)\,\alpha(gh,k,z)^{-1}\alpha(g,h,kz)^{-1}.
\end{equation}
Comparing this with 
the cocycle condition for $\alpha$,
\begin{equation}
\alpha(h,k,z)\alpha(g,hk,z)\alpha(g,h,k)=\alpha(gh,k,z)\alpha(g,h,kz),
\end{equation}
we find that $\omega(g,h,k)=\alpha(g,h,k)$.

\section{The reference category}
\label{app:reference}

In this appendix we construct the reference category $\cR$ used in Sec.~\ref{sec:fermionicII}.
This is a weakly-pointed super fusion category with $S(\cR)\simeq\bZ_2$,
such that the simple object for the nontrivial element of $\bZ_2$
has a Majorana zero mode.\footnote{%
From the discussion of Sec.~\ref{sec:fermionicII},
which requires only the existence of a \emph{single} reference category,
we know that the set of equivalence classes of such categories
corresponds to $N=1,3,5,7\in SSH^3(B\bZ_2)\simeq\bZ_8$.
So, more precisely, we should write $\cR_N$ for $N=1,3,5,7$,
and what we use in Sec.~\ref{sec:fermionicII} is $\cR_1$.
Below, we will construct $\cR_N$ for all four cases
uniformly, and explain how to distinguish them
without depending on the conventions.
}
There are many methods to do this.
As such a category only contains one nontrivial simple object,
it is not too difficult to work out the associators and the pentagon equation directly, in a skeletal form.
We can also start from the Ising category 
and modify it to produce a required super fusion category.
This is not too difficult either.

Here, we provide a rather different construction,
based on the endomorphism category of an infinite-dimensional superalgebra.
This requires much more effort than the other two approaches,
but is in some sense the least computational,
and has some flavor of being based on a physical model,
although we do not elaborate on this point of view here.
This method also allows for a generalization to the super fusion categories over $\bR$,
corresponding to the time-reversal symmetric case, without too much difficulty.
For this extension, see \cite{Kawasumi}.

\subsection{Description of the reference categories $\cR_N$}
\paragraph{The algebra:}
Let $N$ be an odd integer modulo $8$.
Let $C=\Cl(e_1,e_2,\ldots)$ be the complex Clifford algebra generated by infinitely many
anticommuting odd elements $e_k$ with $e_k^2=1$.
It can be thought of as acting on a fermionic half-infinite chain, 
whose $j$-th site has a complex fermion formed by $e_{2j-1}$ and $e_{2j}$.
Let $\rho$ be the shift by one site, so that $\rho(e_k)=e_{k+2}$.
We also let
\begin{equation}
P=\frac12(1-i^Ne_1e_2).
\end{equation}
Note that $i^N=\pm i$ for odd $N$.
Therefore $P$ is a projector onto a state, which we denote by $\ket0$, of the first site.
We let $L$ be the algebra generated by $C$ and two even elements $s$, $t$ subject to
\begin{equation}
ts=1,\qquad st=P,\qquad sa=\rho(a)s\quad (a\in C).
\end{equation}
Pseudo-physically, the operation $s$ inserts a new first site in $\ket0$
and shifts everything to the right.
Conversely, the operation $t$ removes the first site if it is in $\ket0$,
and gives zero otherwise.
We set $s_0=s$, $s_1=e_1s$, $t_0=t$, $t_1=te_1$.
Using $Pe_1P=0$ and $e_1Pe_1=1-P$, we find
\begin{equation}
t_0s_0=t_1s_1=1,\qquad t_0s_1=t_1s_0=0,\qquad s_0t_0+s_1t_1=1.
\label{eq:Leavitt}
\end{equation}

Actually, these are the defining relations of the Leavitt algebra $L(1,2)$ \cite{Leavitt1962,Leavitt1965},
to which $L$ is isomorphic;
see also \cite{AbramsAraSilesMolina}.
The use of endomorphisms of such algebras to construct fusion categories goes back to Izumi \cite{Izumi1}.

\paragraph{The objects $\Pi$ and $M$:}
The shift $\rho$ extends to $L$ by $\rho(x)=s_0xt_0+(-1)^{|x|}s_1xt_1$,
which is an endomorphism thanks to \eqref{eq:Leavitt}.
It satisfies $\rho(x)s_0=s_0x$, $\rho(x)s_1=(-1)^{|x|}s_1x$,
$t_0\rho(x)=xt_0$ and $t_1\rho(x)=(-1)^{|x|}xt_1$,
i.e.~$s_0$ and $s_1$ are an even and an odd element of $\Hom(\id,\rho)$,
and $t_0$ and $t_1$ are an even and an odd element of $\Hom(\rho,\id)$, respectively.
Then \eqref{eq:Leavitt} says that $\rho\simeq 1\oplus\Pi$,
where $1$ is the image of $s_0t_0=P$ and $\Pi$ is the image of $s_1t_1=1-P$,
and $t_1$ restricted to $\Pi$ is an odd isomorphism $\Pi\to 1$.
Furthermore, $\End(\rho)=\Cl(e_1,e_2)$.

\paragraph{The object $\sigma$:}
We let $\sigma$ be the endomorphism of $L$ defined by
\begin{equation}
\sigma(e_k)=e_{k+1},\qquad \sigma(s)=Ws,\qquad \sigma(t)=tW^{-1},
\end{equation}
where
\begin{equation}
W=e^{2\pi iN/8}\cdot\frac12\bigl(1-e_1e_2-e_2e_3+e_1e_3\bigr).
\label{eq:choice2}
\end{equation}
The even element $W$ satisfies 
\begin{equation}
We_1W^{-1}=e_2,\quad We_2W^{-1}=e_3,\quad We_3W^{-1}=e_1
\end{equation}
and commutes with $e_k$ for $k\ge 4$.

Using this, one checks that $\sigma$ preserves the defining relations of $L$:
e.g.~$\sigma(s)\sigma(t)=WPW^{-1}=\frac12(1-i^Ne_2e_3)=\sigma(P)$.
The square $\sigma\circ\sigma$ agrees with $\rho$ on $C$.
A direct computation shows that they also agree on $s$ and $t$, i.e.~$\sigma(W)Ws=\rho(s)$.\footnote{%
This condition requires the phase factor in the definition of  $W$ to be $i^N$,
and therefore allows the coefficient to be $\pm e^{2\pi iN/8}$. Choosing the minus sign
simply sends $N\mapsto N+4$ mod 8, so we can take the plus sign
without loss of generality.
}
Therefore
$
\sigma\otimes\sigma=\rho\simeq 1\oplus\Pi$.

Let us determine $\End(\sigma^n)$.
An element $x\in \End(\sigma^n)$ cannot involve $s$ or $t$,
as they shift all the sites. 
Therefore, $x$ is in $C$ and graded-commutes with $e_k$ for $k>n$, and hence
\begin{equation}
\End(\sigma^n)=\Cl(e_1,\ldots,e_n).
\end{equation}
Since $1\otimes x=\sigma^a(x)$ for $x\in\End(\sigma^b)$,
the tensor product $\End(\sigma^a)\otimes\End(\sigma^b)\to\End(\sigma^{a+b})$ is given by $e_k\otimes 1=e_k$ and $1\otimes e_k=e_{a+k}$.
In particular $\sigma$ is a simple object with a Majorana zero mode $e_1$.
It is self-dual, with the evaluation and coevaluation given by $t_0$ and $s_0$:
indeed, using $PWP=e^{2\pi iN/8}\,\frac{1-i^N}{2}P$, we find
\begin{equation}
t_0\,\sigma(s_0)=\sigma(t_0)\,s_0=\frac{(-1)^{(N^2-1)/8}}{\sqrt2}\neq 0.
\label{eq:FS}
\end{equation}

To summarize, the simple objects are $1$, $\Pi$ and $\sigma$, with $\sigma\otimes\sigma\simeq 1\oplus \Pi$,
and $\sigma\otimes \Pi\simeq\Pi\otimes\sigma\simeq\sigma$.
This is the category $\cR_N$.

\paragraph{The canonical isomorphism $c_{M,\sigma}$:}
Let us take $M=\sigma^2=\rho$.
The canonical isomorphism $c_{M,Y}:M\otimes Y\to Y\otimes M$ is, in the present language,
\begin{equation}
c_{M,Y}=Y(s_0)\,t_0+Y(s_1)\,t_1,
\end{equation}
where the first term corresponds to the summand $1$ and the second to the summand $\Pi$,
for which we used the odd isomorphism $t_1:\Pi\to 1$ and its inverse $s_1$.
For $Y=\sigma$, we have $\sigma(s_0)=Ws$ and $\sigma(s_1)=e_2Ws=We_1s$, and therefore
\begin{equation}
c_{M,\sigma}=WP+We_1Pe_1=W
\label{eq:cMsigmaN}
\end{equation}
for all $N$.
Note that this does not involve any choice of the Majorana operators $\chi$, $\chi'$ on $M$.

\paragraph{Distinguishing $\cR_N$:}
Let us now discuss a convention-independent quantity 
which distinguishes the categories $\cR_N$.
Note that $c_{M,\sigma}$ is an even element of $\End(M\otimes\sigma)$,
which is a Clifford algebra with three generators,
whose even part is a $2\times 2$ matrix algebra.
Therefore there is a standard trace $\mathrm{tr}$ on the even part, with $\mathrm{tr}\,1=2$.
We extend it to the entirety of $\End(M\otimes\sigma)$ 
by setting the odd part to have zero trace.
Then our invariant is $\mathrm{tr}\,c_{M,\sigma}\in\bC$.
Note that this definition does not depend on our specific 
constructions of $\cR_N$,
and applies to any super fusion category, which we assume to be strict,
containing a simple object $\sigma$
with a Majorana zero mode such that $\sigma\otimes\sigma\simeq 1\oplus\Pi$,
where we set $M:=\sigma\otimes\sigma$.

Let us evaluate this invariant for $\cR_N$,
for which  $c_{M,\sigma}=W$.
As the trace of $1$ is $2$ and that of $e_1e_2$, $e_2e_3$ and $e_1e_3$ is zero, we find
\begin{equation}
\mathrm{tr}\,c_{M,\sigma}=e^{2\pi iN/8}.
\label{eq:spinRN}
\end{equation}
Therefore the four categories $\cR_N$ for $N=1,3,5,7$ are distinct.

\paragraph{Spin selection rule:}
This invariant also has a direct physical meaning.
For the bosonic case, the corresponding analysis was given in \cite{Chang:2018iay}.
We will only outline the argument;
a more precise treatment should be possible,
e.g.~by using the tube algebra for the super fusion category.

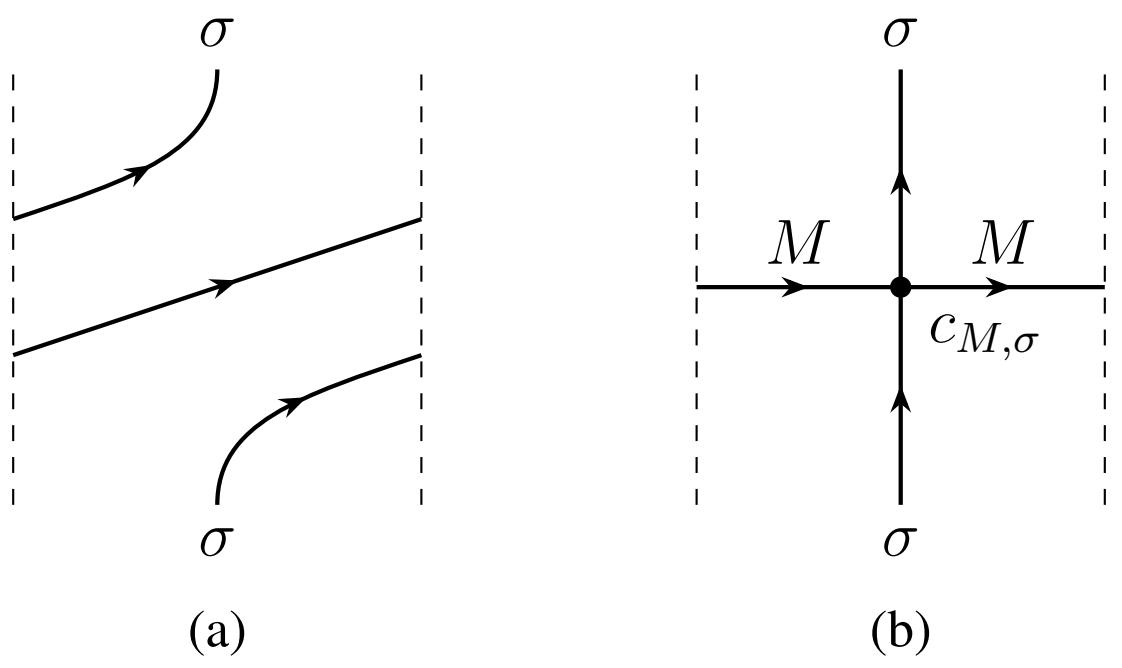
\begin{figure}[h]
    \centering
    \subfloat[\label{fig:winding-a}]{%
    \begin{tikzpicture}[baseline=(o),scale=0.9]
        \coordinate (o) at (0,0);
        \draw[dashed] (-1.5,-1.6) -- (-1.5,1.6);
        \draw[dashed] (1.5,-1.6) -- (1.5,1.6);
        \draw[line] (0,-1.6) node[below] {$\sigma$} .. controls (0,-1) and (0.6,-0.8) .. (1.5,-0.5);
        \draw[line] (-1.5,-0.5) -- (1.5,0.5);
        \draw[line] (-1.5,0.5) .. controls (-0.6,0.8) and (0,1) .. (0,1.6) node[above] {$\sigma$};
    \end{tikzpicture}}
    \hspace{4em}
    \subfloat[\label{fig:winding-b}]{%
    \begin{tikzpicture}[baseline=(o),scale=0.9]
        \coordinate (o) at (0,0);
        \draw[dashed] (-1.5,-1.6) -- (-1.5,1.6);
        \draw[dashed] (1.5,-1.6) -- (1.5,1.6);
        \draw[line] (0,-1.6) node[below] {$\sigma$} -- (o);
        \draw[line] (o) -- (0,1.6) node[above] {$\sigma$};
        \draw[line] (-1.5,0) -- node[above] {$M$} (o);
        \draw[line] (o) -- node[above] {$M$} (1.5,0);
        \node[blob,minimum size=4pt] at (o) {};
        \node[below right=1pt] at (o) {$c_{M,\sigma}$};
    \end{tikzpicture}}
    \caption{(a) The rotation by $4\pi$ makes the line $\sigma$ wind twice around the cylinder.
    The time flows upward, and the two dashed lines are identified.
    (b) The same configuration, regarded as the line $M=\sigma\otimes\sigma$ wrapping around the cylinder
    and crossing $c_{M,\sigma}$.
    }
    \label{fig:winding}
\end{figure}

We put the system on a circle, with the line $\sigma$ inserted along the time direction,
and consider the rotation of the circle by $4\pi$,
which acts on the resulting twisted Hilbert space by $e^{4\pi i(L_0-\bar L_0)}$,
where $L_0-\bar L_0$ is the spin.
This operation makes the line $\sigma$ wind twice around the cylinder,
as shown in Fig.~\ref{fig:winding-a}.
The two strands crossing the dashed line there form $M:=\sigma\otimes\sigma$.
Therefore, the same configuration can be regarded as the line $M$ wrapping around the cylinder and crossing $\sigma$,
see Fig.~\ref{fig:winding-b}.
The crossing there is 
\begin{equation}
c_{M,\sigma}:M\otimes\sigma\to\sigma\otimes M.
\end{equation}

When an element $x\in\End(M\otimes\sigma)$ is inserted on $M$,
this setup  would evaluate to $\mathrm{tr} x$
without the loop of $M$,
for a suitable spin structure around the circle.
Therefore we expect
\begin{equation}
e^{4\pi i(L_0-\bar L_0)}=\mathrm{tr}\, c_{M,\sigma}.
\label{eq:spin}
\end{equation}
The right-hand side is defined purely in terms of the category and its object $\sigma$,
without making any arbitrary choice.

We found above that the right-hand side is $e^{2\pi iN/8}$ for $\cR_N$.
This is exactly the value one would expect 
for the chiral $\bZ_2$ symmetry of $N$ Majorana fermions on the circle.
This allows us to identify $\cR_N$ with the category describing
the chiral $\bZ_2$ symmetry $(-1)^{F_L}$ of $N$ Majorana fermions.

\subsection{Computations of $\nu_{\cR\cR}$ and $\omega_{\cR\cR}$}
\label{app:RR}
Here we derive \eqref{eq:nuomegaRR}.
Throughout this subsection, $\cR$ denotes $\cR_1$, as in Sec.~\ref{sec:fermionicII}.
We use $x,y\in H^1(B(\bZ_2\times\bZ_2);\bZ_2)$ given by $x(a,b)=a$ and $y(a,b)=b$.

\paragraph{Determination of $\nu_{\cR\cR}$:}
$\nu_{\cR\cR}$ is a 2-cocycle of $\bZ_2\times\bZ_2$ with values in $\bZ_2$, defined up to coboundaries,
and $H^2(B(\bZ_2\times\bZ_2);\bZ_2)\simeq \bZ_2^3$ is generated by the classes of
$x\cup x$, $y\cup y$ and $x\cup y$.
If we restrict to a single factor of $\bZ_2$, we get a copy of $\cR$,
for which we already know $\nu=0$.
Therefore, $\nu_{\cR\cR}$ can be taken to be trivial or $x\cup y$.
To determine which, we compare $u_{(1,0),(0,1)}:Y_{1,0}\otimes Y_{0,1}\to Y_{1,1}\otimes M$ and $u_{(0,1),(1,0)}:Y_{0,1}\otimes Y_{1,0}\to Y_{1,1}\otimes M$.
Writing $e$ and $e'$ for the Majorana operators on the first and the second copy of $\sigma$,
both source objects are identified with $\sigma\boxtimes\sigma$ by the interchange law,
which introduces no signs here,
and $\psi_{(1,0)}$, $\psi_{(0,1)}$ become $e\boxtimes 1$ and $1\boxtimes e'$ in both cases.
The difference is that $u_{(1,0),(0,1)}$ maps $(e\boxtimes1,1\boxtimes e')$ to $(\chi,\chi')$,
while $u_{(0,1),(1,0)}$ maps $(1\boxtimes e',e\boxtimes1)$ to $(\chi,\chi')$.
Therefore $u_{(0,1),(1,0)}\circ u_{(1,0),(0,1)}^{-1}$ is an automorphism of $Y_{1,1}\otimes M$ exchanging $\chi$ and $\chi'$.
As $\End(Y_{1,1}\otimes M)=\End(M)=\Cl(\chi,\chi')$,
it is given by an element $w\in\Cl(\chi,\chi')$ with $w\chi=(-1)^{|w|}\chi'w$ and $w\chi'=(-1)^{|w|}\chi w$,
which forces $w\propto\chi-\chi'$.
This is odd, and therefore
\begin{equation}
\nu_{\cR\cR}((1,0),(0,1))+\nu_{\cR\cR}((0,1),(1,0))=1 \mod 2.
\end{equation}
Since $\bZ_2\times\bZ_2$ is abelian, the left-hand side is invariant under the coboundaries $\delta\lambda(g,h)=\lambda(g)+\lambda(h)+\lambda(gh)$.
It is $1$ for $x\cup y$ and $0$ for the trivial cochain,
and therefore we can choose the coordinate system so that $\nu_{\cR\cR}=x\cup y$.

\paragraph{Determination of $\omega_{\cR\cR}$:}
Fix the coordinate system so that $\nu_{\cR\cR}=x\cup y$.
Then $\omega_{\cR\cR}$ and the right-hand side $\omega_0$ of \eqref{eq:nuomegaRR}
both satisfy $\delta\omega=\frac12\nu_{\cR\cR}\cup\nu_{\cR\cR}$,
so $r:=\omega_{\cR\cR}/\omega_0$ is a $U(1)$-valued 3-cocycle,
and it suffices to show that $r$ is a coboundary; we can then absorb it by rescaling $u$.
Now $H^3(B(\bZ_2\times\bZ_2);U(1))\simeq H^4(B(\bZ_2\times\bZ_2);\bZ)\simeq\bZ_2^3$
injects into $H^4(B(\bZ_2\times\bZ_2);\bZ_2)$, with image spanned by $x^4$, $y^4$, $x^2y^2$.
Their restrictions to the subgroups generated by $(1,0)$, $(0,1)$, $(1,1)$
are $(1,0,1)$, $(0,1,1)$, $(0,0,1)$, so a class is detected by these restrictions.
On each such subgroup $\{0,g\}$ the class of $r$ is given by $r(g,g,g)$,
and $\omega(g,g,g)$ is unchanged under the coordinate changes $\lambda$ and $\xi$.
On the two factors we get copies of $\cR$, so $\omega_{\cR\cR}(g,g,g)=1=\omega_0(g,g,g)$.
On the diagonal $\omega_0=-i$, so it remains to show that $\omega_{\cR\cR}=-i$ there.

We realize $\cR\boxtimes\cR$ by the endomorphisms of the graded tensor product of two copies of $L$,
denoting the generators of the second copy by $f_k$, $S$, $T$,
and writing $\sigma_e$, $\sigma_f$, $\rho_e$, $\rho_f$ for $\sigma$ and $\rho$ acting on one copy.
Then $Y_{1,0}=\sigma_e$, $Y_{0,1}=\sigma_f$,
and we take $Y_{1,1}$ to be the image of $Q=\frac12(1-ie_1f_1)$ in $\sigma_e\sigma_f$.

On the diagonal subgroup, $m=0$, so we only need $u:Y_{1,1}\otimes Y_{1,1}\to 1$ and \eqref{eq:omegadef}.
The source is $\rho_e\rho_f$ with the projector $Q\,\sigma_e\sigma_f(Q)=Q\cdot\frac12(1-ie_2f_2)$.
From $Pe_1e_2=iP$ and $(1-P)e_1e_2=-i(1-P)$ we have $te_1e_2=it$ and $te_2=-ite_1$, and similarly for $T$.
Using these, one checks that
\begin{equation}
u=tT(f_1+ie_1)\in\Hom(\rho_e\rho_f,\id)
\end{equation}
satisfies $ue_1f_1=ue_2f_2=iu$, i.e.~$u\,Q\,\sigma_e\sigma_f(Q)=u$.
As $u$ is odd, $\nu_{\cR\cR}((1,1),(1,1))=1$, consistent with $\nu_{\cR\cR}=x\cup y$.

With $g=h=k=(1,1)$, the relation \eqref{eq:omegadef} reads $\sigma_e\sigma_f(u)=\omega\, u$,
where both sides are multiplied from the right by the projector $Q_3:=Q\,\sigma_e\sigma_f(Q)\,\rho_e\rho_f(Q)$ of $Y_{1,1}^{\otimes 3}$.
The identities above give $t\,W^{-1}=t(1-e_1e_3)/\sqrt2$, and therefore
\begin{equation}
\sigma_e\sigma_f(u)=tW_e^{-1}\,TW_f^{-1}(f_2+ie_2)
=-\frac{i}2\,u\,(1-e_1e_3)(1-f_1f_3).
\end{equation}
In the product $(1-e_1e_3)(1-f_1f_3)Q_3$, the terms linear in $e_1e_3$ or $f_1f_3$ vanish after multiplying $u$ from the left,
since $e_1e_3$ and $f_1f_3$ anticommute with $e_1f_1$ and $uQ=u$.
In the last term $e_1e_3f_1f_3=-(e_1f_1)(e_3f_3)$ we can use $ue_1f_1=iu$ and $e_3f_3\,\rho_e\rho_f(Q)=i\rho_e\rho_f(Q)$.
Therefore $\sigma_e\sigma_f(u)Q_3=-i\,uQ_3$, i.e.
\begin{equation}
\omega_{\cR\cR}((1,1),(1,1),(1,1))=-i.
\end{equation}

\bibliographystyle{ytamsalpha}
\def\arxivfont{\rm}
\baselineskip=.93\baselineskip
\bibliography{ref}

\end{document}